\documentclass[useAMS,usenatbib]{mn2e}

\usepackage{graphicx,amssymb,amsmath,times}
\usepackage{natbib,graphicx}
\usepackage[a4paper]{hyperref}
\usepackage{lineno}
\usepackage{deluxetable}

\usepackage{Times}

\newcommand{\arcm}{\ensuremath{\mathrm{^\prime}}}

\newcommand{\planck}{\textit{Planck}}
\newcommand{\swift}{\textit{Swift}}
\newcommand{\fermi}{\textit{Fermi}}
\newcommand{\chandra}{\textit{Chandra}}

\newcommand{\quattroc}{4C~+49.22}

\vspace{-1.0cm}

\title[\fermi, \swift, \planck\ view of 4C~+49.22 (S4~1150+49)]
{Radio-Gamma-ray connection and spectral evolution in 4C~+49.22 (S4~1150+49): the \fermi, \swift\ and \planck\ view}
\author[S. Cutini et~al.]
{
S. Cutini$^{ 1,2}$\thanks{E-mail:sara.cutini@asdc.asi.it},
S. Ciprini$^{1,2}$,
M. Orienti$^{3,4}$,
A. Tramacere$^{5}$,
F. D'Ammando$^{3,6,7}$,
\and
F. Verrecchia$^{1,2}$
G. Polenta$^{1,2}$,
L. Carrasco$^{8}$,
V. D'Elia$^{1,2}$,
P. Giommi$^{1}$,
\and
J.~Gonz\'{a}lez-Nuevo$^{9,10}$,
P. Grandi$^{11}$,
D.~Harrison$^{12,13}$,
E. Hays$^{14}$,
S. Larsson$^{15,16,17}$,
\and
A.~L\"{a}hteenm\"{a}ki$^{19}$,
J.~Le\'{o}n-Tavares$^{18,21}$,
M.~L\'{o}pez-Caniego$^{9}$,
P.~Natoli$^{1,22}$,
R. Ojha$^{14}$,
\and
B.~Partridge$^{23}$,
A. Porras$^{8}$,
L. Reyes$^{20}$,
E. Recillas$^{8}$,
\and
E. Torresi$^{11}$
\\ $~$  \\
$^{1}${ASI Science Data Center, Via del Politecnico snc, I-00133 Rome, Italy}\\
$^{2}${INAF Osservatorio Astronomico di Roma, Monte Porzio Catone ,Via Frascati 33, I-00040 Rome, Italy}\\
$^{3}${INAF Istituto di Radioastronomia, I-40129 Bologna, Italy}\\
$^{4}${Dipartimento di Astronomia, Universita' di Bologna, via Ranzani 1, I-40127 Bologna, Italy}\\
$^{5}${ISDC Data Centre for Astrophysics, University of Geneva, Ch. d'Ecogia 16, Versoix, Switzerland}\\
$^{6}${Dipartimento di Fisica, Universit\'a degli Studi di Perugia, Via A. Pascoli, I-06123 Perugia, Italy}\\
$^{7}${INFN Sezione Perugia, Via A. Pascoli, I-06123, Perugia, Italy}\\
$^{8}${Instituto Nacional de Astrofisica, Optica y Electronica, Tonantzintla, Puebla, C.P 72860 Mexico}\\
$^{9}${Instituto de F\'{\i}sica de Cantabria (CSIC-Universidad de Cantabria), Avda. de los Castros s/n, Santander, Spain}\\
$^{10}${SISSA, Astrophysics Sector, via Bonomea 265, I-34136 Trieste, Italy}\\
$^{11}${Istituto di Astrofisica Spaziale e Fisica Cosmica-Bologna, INAF, via Gobetti 101, I-40129 Bologna, Italy}\\
$^{12}${Institute of Astronomy, University of Cambridge, Madingley Road, Cambridge CB3 0HA, U.K.}\\
$^{13}${Kavli Institute for Cosmology Cambridge, Madingley Road, Cambridge, CB3 0HA, U.K.}\\
$^{14}${ORAU/NASA Goddard Space Flight Cetner, Greenbelt, MD 20771, USA}\\
$^{15}${Department of Physics, Stockholm University, AlbaNova, SE-106 91 Stockholm, Sweden}\\
$^{16}${The Oskar Klein Centre for Cosmoparticle Physics, AlbaNova, SE-106 91 Stockholm, Sweden}\\
$^{17}${Department of Astronomy, Stockholm University, SE-106 91 Stockholm, Sweden}\\
$^{18}${Instituto Nacional de Astrof\'{\i}sica \'Optica y Electr\'onica (INAOE), Apartado Postal 51 y 216, 72000 Puebla, M\'exico}\\
$^{19}${Helsinki Institute of Physics, Gustaf H\"{a}llstr\"{o}min katu 2, University of Helsinki, Helsinki, Finland}\\
$^{20}${Physics Department, California Polytechnic State University, San Luis Obispo, CA 94307, USA}\\
$^{21}${Finnish Centre for  Astronomy with ESO (FINCA), University of Turku,V\"ais\"al\"antie 20, FI-21500  Piikki\"o, Finland}\\
$^{22}${Dipartimento di Fisica e Scienze della Terra, Universit\`a degli Studi di Ferrara e Sezione INFN di Ferrara, via Saragat 1, I-44100 Ferrara, Italy}\\
$^{23}${Haverford College Astronomy Department, 370 Lancaster Avenue, Haverford, Pennsylvania, U.S.A.}\\
\\$^{\bigstar}${Correspondence: sara.cutini@asdc.asi.it}\\
\vspace{-1.0cm}
} 
\begin{document}
\date{Accepted 2014 ... .. Received 2014 ... .. ; in original form 2013 ... ..}
\pagerange{\pageref{firstpage}--\pageref{lastpage}} \pubyear{2002}
\maketitle
\label{firstpage}
%
\begin{abstract}
%

The Large Area Telescope on board the \textit{Fermi} \textit{Gamma-ray} \textit{Space} \textit{Telescope} detected a strong $\gamma$-ray flare on 2011 May 15 from a source identified  as \quattroc, a flat spectrum radio quasar also known as S4~1150+49. This blazar, characterised by a prominent radio-optical-X-ray jet, was in a low $\gamma$-ray activity state during the first years of \fermi\ observations. Simultaneous  observations during the quiescent, outburst and post-flare $\gamma$-ray states were obtained by  \swift, \planck\ and optical-IR-radio telescopes (INAOE, Catalina CSS, VLBA, Mets\"{a}hovi). The flare is observed from microwave to X-ray bands with correlated variability and the \fermi, \swift\ and \planck\ data for this FSRQ  show some features more typical of BL Lac objects, like the synchrotron peak in the optical band that outshines the thermal blue-bump emission, and the X-ray spectral softening. Multi-epoch VLBA observations show the ejection of a new component close in time with the GeV $\gamma$-ray flare. The radio-to-gamma-ray spectral energy distribution is modeled and fitted successfully for the outburst and the post-flare epochs using either a single flaring blob with two emission processes (synchrotron self Compton,  and external-radiation Compton), and a two-zone model with SSC-only mechanism.

$~~$
\end{abstract}


%
\section{Introduction}\label{sect:introduction}
%

Data and results from simultaneous and coordinated $\gamma$-ray and multi-wavelength (MW) observations of the flat spectrum radio quasar (FSRQ) \quattroc\ (also known as S4~1150+49, OM~484, SBS~1150+497 and GB1 1150+497), are presented.



\quattroc\ is a core-dominated, radio-loud  FSRQ located at $z = 0.334$ \citep{lynds68,burbidge68,stepanian01}. The Sloan  Digital Sky Survey \citep[SDSS;][]{adelman} DR7 and DR8 give values of $z=0.3339$ and of $z=0.33364$, respectively. This blazar shows a kiloparsec-extent and one-sided, knotty and wiggling radio jet, with high surface brightness, sharp bends and resolved substructures \citep[see, e.g., ][]{owen84,akujor91,sambruna04,sambruna06a,sambruna06b}. The jet, known to show constant low optical polarisation \citep{moore81}, has a twisted morphology with a corkscrew structure reminiscent of 3C 273 and, remarkably, is also well detected at X-ray and optical bands. The 10$\arcsec$ X-ray jet of \quattroc\, is one of the brightest known among blazars, and is an example of X-ray emission produced by inverse Compton (IC) scattering of the cosmic microwave background (CMB) photons \citep{tavecchio05,hardcastle06,sambruna06a}.

\par The \chandra\ \textit{X-ray} \textit{Observatory} detected a Fe K-shell emission line in \quattroc\, consistent with fluorescent $K\alpha$ emission from cold iron \citep{gambill03,sambruna06a,sambruna06b}. The estimated mass of the super-massive black hole (SMBH) is $3.3\times 10^{8} M_{\odot}$ according to the FWHM of the broad $H\beta$ line \citep[4810 km s$^{-1}$;][]{shields03} and is $1.6\times 10^{9} M_{\odot}$ according to the estimation from the host galaxy luminosity \citep{decarli08}. From the SDSS R5 spectrum the continuum luminosity of the Broad Line Region (BLR) at 5100 \AA~  is evaluated to be $\ F_{\lambda}= 10^{44.6}$ erg s$^{-1}$ with a BLR size of $R_{BLR}=1.26\times 10^{17}$ cm \citep{decarli08}.

\par \quattroc\ showed a large $\gamma$-ray outburst detected with the  Large Area Telescope  onboard  the  \textit{Fermi} \textit{Gamma-ray} \textit{Space} \textit{Telescope} (\fermi-LAT), at energies above 100 MeV on  2011 May 15 \citep{reyes11}. Before this flaring event the source was in a long-standing quiescent state with no detection reported in the first \fermi-LAT source catalogue \citep{fermi_1fgl_catalog} or in previous source catalogs released by other MeV-GeV $\gamma$-ray missions like EGRET \citep[][]{hartman} and AGILE \citep[][]{pittori}. It was included in the second \fermi-LAT  source catalogue \citep[2FGL hereafter; 2FGL J1153.2+4935]{fermi_2fgl_catalog} with a 2-year averaged $\gamma$-ray flux ($E>100$ MeV) of ($2.6 \pm 0.4$) $\times10^{-8}$~photons~cm$^{-2}$~s$^{-1}$.
During the flare the source reached a flux almost two orders of magnitude higher than the 2FGL average flux. The source was significantly detected on a daily timescale  in 2011 April \citep{hays11}. The  $\gamma$-ray spectrum did not show significant changes during the outburst compared to the pre and post-flare days. In the post-flare phase \swift-XRT reported an X-ray flux  six times higher than previous archival XRT observations \citep{reyes11}. This FSRQ was also  observed with the \planck\ satellite. According to the \planck-On-the-{Fly} Forecaster, \quattroc\ was observed by \planck\ from 2011 May 11 to May 26. We exploited \planck, \swift\ and \fermi\ simultaneous data, for the first time, to study this  blazar.
We collected Spectral Energy Distribution (SED) archival data from several surveys and telescopes, from radio to $\gamma$ rays: Dixon Master List of Radio Sources \citep{dixon95}; the FIRST Survey Catalog of 1.4-GHz Radio Sources \citep{becker12}; Kuehr Extragalactic Radio Sources at 5 GHz \citep{kuehr}; the NRAO VLA Sky Survey  \citep{nvss}; the VLA Low-Frequency Sky Survey at 74 MHz \citep{vlss}; the Green Bank 6-cm  Catalog of Radio Sources \citep{gb6}; the 20-cm Northern Sky Catalog;  the {\it Planck} Early Release Compact Source Catalogue \citep{planck2011-1.10}; Five-Year Wilkinson Microwave Anisotropy Probe \citep{WMAP}; the SDSS; the {\it ROSAT} All-Sky Survey Bright Source Catalogue \citep{rass}, the {\it ROSAT} Catalog of PSPC WGA Sources \citep{wga}; and the  2FGL catalog.
\par The paper is organised as follows. In Section \ref{sect:LAT} the \fermi-LAT data analysis is presented, while in Section \ref{sect:planck} the millimetre \planck\ data from the sky coverages with Planck are described, with particular attention to the fourth sky scan that is coincident in time with the 2011 May  $\gamma$-ray outburst. In Section \ref{sect:swift} optical, UV and X-ray data from nine \swift\ pointings under the Target of Opportunity programme (ToO) performed between 2011 April 26 and May 25 are presented. Section \ref{sect:radio-optical} reports on ground-based radio-to-optical observations obtained by the VLBA (MOJAVE monitoring programme) and Mets\"{a}hovi radio observatories, and by near-IR and optical photometric observations of the INAOE observatory and the Catalina Sky Survey (CSS). Section \ref{sect:gamma-lightcurve} characterises the $\gamma$-ray variability and cross correlations in \quattroc\ through 3 years of \fermi-LAT survey data, with particular focus on the outburst of 2011 May. Multi-epoch SEDs are built for the source and modeled with  a one-zone with two components Synchrotron Self-Compton (SSC) and  External Radiation Compton (ERC) description and a two-zone Synchrotron Self-Compton description in Section \ref{sect:SED}. This allows us to infer both the production sites of the high-energy emission and emission scenarios. A summary and conclusions are reported in Section \ref{sec:discussion}.
We adopted a standard spatially-flat six-parameter $\Lambda$ cold dark matter ($\Lambda$-CDM) cosmology based on \textit{Planck} results \citep{plank_cosmological_parameters_paper}, namely  with $\Omega_{m}$=0.315 and $H_0$= 67.3 km s$^{-1}$ Mpc$^{-1}$. The corresponding luminosity distance at $z=0.334$  is $d_L$= 1822.3 Mpc.

%
%

%
%

\section{$\gamma$-ray observations and analysis of FERMI-LAT data}\label{sect:LAT}
%
%
%
%
The LAT instrument is a pair conversion telescope comprising a modular array of 16
towers---each with a tracker module of silicon micro-strip detectors and a hodoscopic calorimeter of CsI(Tl) crystals---surrounded by an  Anti-Coincidence Detector made of tiles of plastic scintillator.
The LAT is capable of measuring the directions and energies of  $\gamma$-ray photons with energies from 20 MeV to $>300$ GeV \citep[for details, see,  ][]{michelsonLAT08,lat-calibration,ackermann2012sys}.
\par
The data presented in this paper were collected in the first three years of \fermi\ science observations, from 2008 August 4 to 2011 August 4  (MJD 54682-55778) with E $>$ 100 MeV.
Photon events were selected using the Pass 7 event classification and reconstruction and the corresponding Instrument Response Functions (IRFs) \texttt{P7SOURCE\_V6}.
This selection provides a clean set of events (in terms of direction, energy reconstruction and background rejection) a large effective area and well understood response functions for point source analysis.
To minimise contamination  from photons produced by cosmic rays interacting with the Earth's atmosphere, $\gamma$-ray events
that have reconstructed directions with angles $>100^{\circ}$ with respect to the
local zenith have been excluded and the time intervals when the rocking angle of the LAT was greater than $52^{\circ}$ were rejected.
\par
The reduction and analysis of LAT data was performed using the \texttt{ScienceTools} v09r23p01\footnote{For  documentation of the Science Tools, see http://fermi.gsfc.nasa.gov/ssc/data/analysis/documentation/.}, specifically using an unbinned maximum-likelihood estimator of the spectral model parameters (\texttt{gtlike} tool).
For \quattroc, which is located at high Galactic latitude, events are
extracted within a $10^\circ$ radius of the region of interest (ROI) centred at the position of the radio source counterpart.
This angular radius, comparable to the 68\% containment
angle of the Point Spread Function (PSF)\footnote{http://www.slac.stanford.edu/exp/glast/groups/canda/lat$\_$Performance.htm.} at the lowest energies, provides sufficient events
to accurately constrain the diffuse emission components.
Following the 2FGL catalogue the spectral model used for \quattroc\ is the power-law flux density distribution of the form $F(E) = N_0(E/E_0)^{-\Gamma}$.
The source region model includes  all point sources in the 2FGL within $20^{\circ}$ of \quattroc\ (source region)  including \quattroc\ itself. The sources within the $10^{\circ}$ radius of ROI were fitted with a power-law flux density distribution with photon indices $\Gamma$ frozen to the values obtained from the likelihood analysis of the full data set, while those beyond $10^{\circ}$ ROI radius had both index $\Gamma$ and normalisation frozen to those found in the 2FGL catalogue.

A Galactic diffuse emission model  (\texttt{gal\_2yearp7v6\_v0.fits}) and Isotropic component (\texttt{iso\_p7v6source.txt}) were used to model the background\footnote{Details on  the background model are available from the {\it Fermi} Science Support Center, see: http://fermi.gsfc.nasa.gov/ssc/data/access/lat/BackgroundModels.html}.

For the light curve extraction, which is presented in Section \ref{sect:gamma-lightcurve}, the Upper Limits (UL) at 2-$\sigma$ confidence level  were computed for time intervals  in which the likelihood Test Statistic  \cite[TS; ][]{mattox96}  was less than  9 or the number of model predicted $\gamma$ rays for \quattroc\ $N_{pred} < 3$ or $\Delta F(E)/ F(E) > 0.5$. The UL estimation  procedures  are described in the 2FGL catalogue paper \citep{fermi_2fgl_catalog}.

Details on the unbinned likelihood spectra  fit for \quattroc\ in the 0.1-100 GeV range are reported in Table \ref{tab:spectralfit} and SED data points for both epochs are reported in Table \ref{tab:sedfermi}.
The estimated systematic uncertainty of the integral fluxes above 100 MeV is about 8.1\%  and -6.9\% for a soft source like \quattroc\ \citep{ackermann2012sys};  the stated uncertainties in the fluxes  are statistical only.



\begin{table*}
\vspace{-0.1cm}%
\caption[]{Summary of the unbinned likelihood spectral
fit above 100 MeV}\label{tab:spectralfit} 
%
\begin{center}
\begin{tabular}{ll} %
\multicolumn{2}{l}{   } \\ \hline%
Interval  & Best-fit Model and Parameters \\  \hline %
%
%
Integrated data   & Power-law  \\ %
2008-08-08  (MJD: 54686)      & $\Gamma= 2.26 \pm 0.04$    \\ %
2011-08-04 (MJD:  55777)     & F$_{E>100~\mathrm{MeV}}$ = (5.8$\pm$ 0.3)$\times 10^{-8}$ [ph cm$^{-2}$ s$^{-1}$]   \\

\hline 
Outburst/high state        &  Power-law  \\ %
2011-05-14 (MJD: 55695)        & $\Gamma= 2.20 \pm 0.09$    \\ %
2011-05-16 (MJD: 55697)       & F$_{E>100~\mathrm{MeV}}$ = (1.5$\pm$ 0.2)$\times 10^{-6}$ [ph cm$^{-2}$ s$^{-1}$]  \\%
\hline 
Post-flare/lower state       &   Power-law  \\ %
2011-05-17 (MJD: 55698)        & $\Gamma= 2.23 \pm 0.08$    \\ %
2011-05-26 (MJD: 55707)         & F$_{E>100~\mathrm{MeV}}$ = (6.4$\pm$ 0.6)$\times 10^{-7}$ [ph cm$^{-2}$ s$^{-1}$]  \\%
%
\hline
\end{tabular}
\end{center}
\end{table*}
%

\section{Simultaneous mm observations and results by \planck}\label{sect:planck}
%
%

\planck\ \citep{tauber2010a,planck2011-1.1, instr13} is the third generation space mission
to measure the anisotropy of the cosmic microwave background.  It observes the
sky in nine frequency bands covering 30--857\,GHz with high sensitivity
and angular resolution from 31\arcm\ to 5\arcm. Full sky coverage is attained in
$\sim$ 7 months. The Low Frequency Instrument (LFI; \citealt{mandolesi2010, planck2011-1.6,lfi13}) covers the 30, 44 and 70\,GHz bands with
amplifiers cooled to 20\,\hbox{K}.  The High Frequency Instrument (HFI;
\citealt{planck2011-1.5,hfi13}) covers the 100, 143, 217, 353, 545  and
857\,GHz bands with bolometers cooled to 0.1\,\hbox{K}.  Polarisation is measured in
all but the highest two bands
\citep{leahy2010, rosset2010}.  A combination of radiative cooling and three
mechanical coolers produces the temperatures needed for the detectors and optics
\citep{planck2011-1.3}.  Two Data Processing Centers (DPCs) check and calibrate the
data and make maps of the sky \citep{planck2011-1.7, planck2011-1.6}.  \planck's
sensitivity, angular resolution, and frequency coverage make it a powerful
instrument for Galactic and extragalactic astrophysics as well as cosmology. The \planck\ beams scan the entire sky exactly twice in
one year, but scan only about 95 \% of the sky in six months.
For convenience, we call an approximately six month period a
``survey'', and use it as a shorthand for one coverage of the sky.
In order to take advantage of the simultaneity between the \planck\ observations and the \fermi-LAT $\gamma$-ray flare of 2011 May 15,
flux densities have been extracted from maps produced using only data collected during a portion of the \planck\ fourth sky survey (2011 May 11-26).
Moreover, for comparison we have also extracted  flux densities from separate maps for the first (2009 November 16-26),
the second (2010 May 11-26) and the third \planck\ survey (2010 November 16-26). Results are reported in Table \ref{tab:planckdata}.
The Early Release Compact Source catalogue (ERCSC, \citealt{planck2011-1.10}) and the \planck\ catalogue of Compact Sources (PCCS, \citealt{planck2013}) include average flux densities for \quattroc.
All of these maps have been produced through the standard LFI and HFI pipelines adopted for the internal DX8 release.
LFI flux densities were obtained at 30, 44 and 70\,GHz using the IFCAMEX code, an
implementation of the Mexican Hat Wavelet 2 (MHW2) algorithm that is
being used in the LFI DPC infrastructure to detect and extract flux
densities of point-like sources in CMB maps. This wavelet is defined as
the fourth derivative of the two-dimensional Gaussian function, where
the scale of the filter is optimized to look for the maximum in the S/N
of the sources in the filtered map, and has been previously applied to
{\it WMAP} and \planck\ data and simulations \citep{gonzaleznuevo06,lopezcaniego06,lopezcaniego07,massardi09}.
First, we obtained a flat patch centred on the source and applied the MHW2
software. This algorithm produces an unbiased estimation of the flux
density of the source and its error. Second, we convert the peak flux
density from temperature units to Jy/sr and then to Jy by multiplying it by
the area of the instrument beam,  taking the beam solid
angle into account.
In this analysis we used the effective Gaussian full width at
half maximum (FWHM) whose area is that of the actual elliptical beam at
30, 44 and 70\,GHz, respectively, as provided by the LFI DPC.
HFI flux densities have been extracted using aperture photometry.
Flux densities were evaluated assuming a circularly symmetric
Gaussian beam of the given FWHM.  An aperture is centred on the position of
the source and an annulus around this aperture is used to evaluate the
background. A correction factor which accounts for the flux of the source
in the annulus may be calculated and is given below, where $k_0$, $k_1$
and $k_2$ are the number of FWHMs of the radius of the aperture, the inner
radius of the annulus and the outer radius of the annulus respectively.
\begin{equation}
\tiny
F_{true}=\left( 1- \left(\frac{1}{2}\right)^{4k_0^2} - \left( \left(\frac{1}{2}\right)^{4k_1^2} -\left(\frac{1}{2}\right)^{4k_2^2} \right) \frac{k_0^2}{k_2^2-k_1^2}\right)^{-1} F_{obs}
\end{equation}
\normalsize
Here we used a radius of 1 FWHM for the aperture, $k_0 =1$; the
annulus is located immediately outside of the aperture and has a width of
1 FWHM, $k_1=1$ and $k_2=2$. The flux density, $F_{true}$ may then be
evaluated from the observed flux density, $F_{obs}$, where $F_{obs}$ is the
total flux inside the aperture after the background has been subtracted.
\planck\ data for each survey are reported in Table~\ref{tab:planckdata}.

\begin{table}
\caption[]{\fermi-LAT spectral energy distribution data points}
\begin{center}
\label{tab:sedfermi}
\begin{tabular}{lll}
\hline
Epoch          &  frequency   &  $\nu~f(\nu)$    \\

              &   [Hz]       & [erg cm$^{-2}$ s$^{-1}$]     \\
\hline
2011-05-15      & (6.0$\pm$0.3)$\times 10^{22}$ & (2.5$\pm$0.3)$\times 10^{-10}$\\
(MJD: 55696)   & (2.4$\pm$0.3)$\times 10^{23}$ & (2.2$\pm$0.3)$\times 10^{-10}$\\
                & (9.5$\pm$0.3)$\times 10^{23}$ & (1.4$\pm$0.4)$\times 10^{-10}$\\
                & (3.8$\pm$0.3)$\times 10^{24}$ & (8.3$\pm$5.9)$\times 10^{-11}$\\
\hline
2011-05-17/25  &     (6.0$\pm$0.3)$\times 10^{22}$ & (1.1$\pm$0.1)$\times 10^{-10}$\\
(MJD: 556978/55706)  &     (2.4$\pm$0.3)$\times 10^{23}$ & (8.6$\pm$1.3)$\times 10^{-11}$\\
               &     (9.5$\pm$0.3)$\times 10^{23}$ & (6.4$\pm$1.7)$\times 10^{-11}$\\
               &     (3.8$\pm$0.3)$\times 10^{24}$  & (4.9$\pm$2.8)$\times 10^{-11}$\\
\hline
\end{tabular}
\end{center}
\end{table}

\begin{table}
\caption[]{Flux densities for \quattroc\ from the four \planck\ surveys. First survey: 2009 November 16-26, second survey: 2010 May 11-26, third survey:  2010 November 16-26, fourth survey: 2011 May 11-26.} 
\begin{tabular}{cccc}
\hline
\planck-LFI & $\nu$  & Flux Density & Errors    \\

survey & [GHz] & [Jy] & [Jy]     \\
\hline

 1$^{st}$& 30&  1.14&   0.15 \\
 2$^{nd}$& 30&  1.66&  0.12 \\
 3$^{rd}$& 30&  1.75&  0.14 \\
 4$^{th}$& 30&  1.46&  0.13 \\
 1$^{st}$& 44&  1.24&   0.26 \\
 2$^{nd}$& 44&  1.93&   0.24 \\
 3$^{rd}$& 44&  1.28&   0.26 \\
 4$^{th}$& 44&  1.95&   0.13 \\
 1$^{st}$& 70&  1.21&   0.13 \\
 2$^{nd}$& 70&  1.85&  0.18 \\
 3$^{rd}$& 70&  1.40&   0.23 \\
 4$^{th}$& 70&  2.36&  0.16 \\
\hline
\planck-HFI & $\nu$  & Flux Density & Errors     \\

survey & [GHz] & [Jy] & [Jy]     \\
\hline

1$^{st}$ & 100 & 0.57 & 0.23 \\
2$^{nd}$ & 100 & 1.93& 0.15  \\
3$^{rd}$ & 100 & 1.58& 0.17  \\
4$^{th}$ & 100 & 2.37& 0.17 \\
1$^{st}$ & 143 & 0.51 & 0.20 \\
2$^{nd}$ & 143 & 1.62& 0.16 \\
3$^{rd}$ & 143 & 1.58& 0.18  \\
4$^{th}$ & 143 & 2.26& 0.15  \\
1$^{st}$ & 217 & 0.41 & 0.10 \\
2$^{nd}$ & 217 & 1.48& 0.11  \\
3$^{rd}$ & 217 & 1.30& 0.08  \\
4$^{th}$ & 217 & 1.98& 0.09  \\
1$^{st}$ & 353 & 0.42& 0.20  \\
2$^{nd}$ & 353 & 0.98 & 0.09  \\
3$^{rd}$ & 353 & 1.04& 0.12  \\
4$^{th}$ & 353 & 1.61& 0.09 \\
1$^{st}$ & 545 & 0.20 & 0.21  \\
2$^{nd}$ & 545 & 0.76 & 0.17  \\
3$^{rd}$ & 545 & 0.72 & 0.18  \\
4$^{th}$ & 545 & 1.34& 0.20  \\
1$^{st}$ & 857 &  0.25 &  UL\\
2$^{nd}$ & 857 & 0.17& 0.23  \\
3$^{rd}$ & 857 & 0.32 & 0.27  \\
4$^{th}$ & 857 & 0.74 & 0.20  \\
\hline

\label{tab:planckdata}
\end{tabular}
\end{table}

\begin{figure}
\hspace{-1.1cm}
 \includegraphics[width=7cm,angle=270]{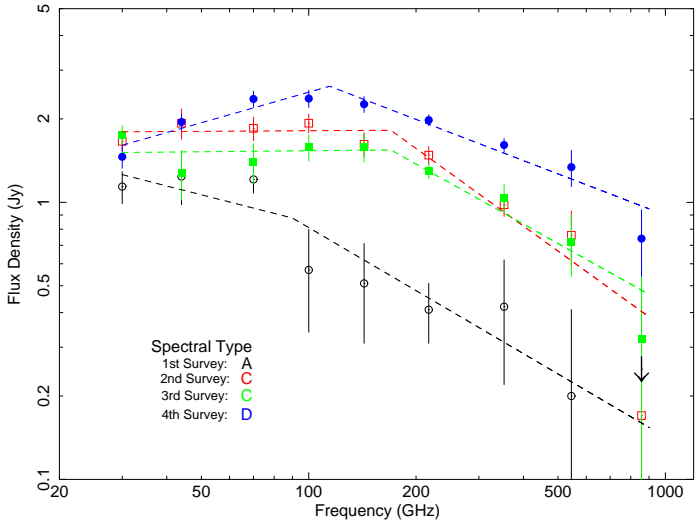}
 \caption{\planck\ flux densities for \quattroc\ at all nine frequencies  as reported in Table \ref{tab:planckdata}. Each frequency is shown in four colours corresponding to four surveys: black open circles correspond to the first survey, red open squares to the second, green filled squares to the third and blue filled circle to the fourth.}
  \label{fig:planckflux}
\end{figure}

Spectra  presented in Figure \ref{fig:planckflux} are modeled with  a  broken power-law; see Equation 5 of  \citet{leon2012}, in order to characterise the evolution of the sub-mm spectra in terms of the  spectral classification presented in  Figure 3 of  \citet{leon2012}. Within the  mentioned classification scheme  the  hard sub-mm spectrum observed during 2009 November  (open circles) can be classified as spectral-type A (both power-law indices, $\alpha_{mm}< 0 $ , $\alpha_{sub-mm} < 0$  and the relative difference between indices is less than 50\%),  thus indicating  the absence  of a new jet component in a very early development stage.
As the plasma blob  propagates down the jet, the shape of the spectrum changes; the relative difference between  $\alpha_{mm}$ and $\alpha_{sub-mm}$  became greater than 50\% as can be seen in 2010 May   and 2010 November, respectively. These spectral shapes  can be classified  as spectral-type C.
As the plasma blob  propagates down the jet, its spectral turnover  shifts to lower frequencies, from 170 GHz of 2010 May and 2010 November to around 100 GHz in 2011 May.
%
%
The sub-mm spectrum of  2011 May, simultaneous to the $\gamma$-ray flare,  shows a well-defined synchrotron component and it is consistent with spectral-type D ($\alpha_{mm}> 0 $ and $\alpha_{sub-mm} < 0$).
Sources with sub-mm spectra classified as spectral type C or D are more likely to be strong $\gamma$-ray emitters, which  is in good agreement  with the fact that \quattroc\ became a $\gamma$-ray emitter  only  after  its sub-mm spectral shape changed to spectral-type D \citep{leon2012}. This spectral shape can be associated with a single synchrotron component that becomes self absorbed in the middle of the mm wavelength regime. Such high spectral turnover frequencies reveal the presence of emerging disturbances in the jet that are likely to be responsible for the high levels of $\gamma$-ray emission \citep{marscher06,marscher2014}.

%
\section{Simultaneous X-ray and UV-optical observations and results from \swift}\label{sect:swift}
%
%
\begin{table*}
\caption{Log and fitting results of the data obtained by
the XRT instrument on board \swift.}
\begin{center}
\begin{tabular}{ccccc}
\hline
Date& Exp. Time & Photon Index & Unabsorbed Flux 0.3--10 keV & $\chi2_{\rm red}$ (d.o.f.)\\
    & (s)     &            & erg cm$^{-2}$ s$^{-1}$ \\
\hline
2008-04-28 (MJD: 54584) & 5438 & 1.82 $\pm$ 0.11 & (4.8 $\pm$ 0.4)$\times$10$^{-12}$ & 0.83 (24) \\
2009-05-06 (MJD: 54957) & 3122 & 1.80 $\pm$ 0.16 & (3.8 $\pm$ 0.5)$\times$10$^{-12}$ & 0.64 (10) \\
2009-11-17 (MJD: 55152) & 5218 & 2.05 $\pm$ 0.14 & (3.8 $\pm$ 0.4)$\times$10$^{-12}$ & 0.71 (17) \\
2011-04-26 (MJD: 55677) & 4647 & 1.99 $\pm$ 0.08 & (1.1 $\pm$ 0.1)$\times$10$^{-11}$ & 0.75
(44) \\
2011-04-29 (MJD: 55680) & 4813 & 1.69 $\pm$ 0.10 & (7.8 $\pm$ 0.6)$\times$10$^{-12}$ & 0.91
(32) \\
2011-05-02 (MJD: 55683) & 4396 & 1.77 $\pm$ 0.11 & (7.6 $\pm$ 0.6)$\times$10$^{-12}$ & 1.10
(27) \\
2011-05-15 (MJD: 55696) & 3611 & 2.03 $\pm$ 0.06 & (2.1 $\pm$ 0.1)$\times$10$^{-11}$ & 0.79
(77) \\
2011-05-17 (MJD: 55698) & 3406 & 1.70 $\pm$ 0.08 & (1.2 $\pm$ 0.1)$\times$10$^{-11}$ & 0.75
(36) \\
2011-05-19 (MJD: 55700) & 3708 & 1.87 $\pm$ 0.08 & (1.2 $\pm$ 0.1)$\times$10$^{-11}$ & 0.92
(42) \\
2011-05-22 (MJD: 55703) & 4141 & 1.74 $\pm$ 0.09 & (8.4 $\pm$ 0.6)$\times$10$^{-12}$ & 0.73
(32) \\
2011-05-23 (MJD: 55704)& 4002 & 1.69 $\pm$ 0.09 & (9.3 $\pm$ 0.7)$\times$10$^{-12}$ & 0.86
(31) \\
2011-05-25 (MJD: 55706) & 3920 & 1.70 $\pm$ 0.10 & (8.8 $\pm$ 0.7)$\times$10$^{-12}$ & 1.03
(31) \\
\hline
2011-05-17/25 (MJD: 55698/55706) & 19176 & 1.76 $\pm$ 0.04 & (1.1 $\pm$ 0.1)$\times$10$^{-11}$ & 0.93
(152) \\
\hline
\end{tabular}
\end{center}
\label{swifttab}
\end{table*}

In response to the high $\gamma$-ray activity of \quattroc,\ the \swift\ satellite \citep{gehrels04} performed 9 ToO observations, between  2011 April 26 and May 25. In order
to reference the source's past activity, we also
analysed the observations performed on 2008 April 8, 2009  May 6, and 2009 May 17.
The 2011 observations were performed using two of three on-board instruments: the X-ray Telescope \citep[XRT; ][0.2--10.0 keV]{burrows05} and the UltraViolet Optical Telescope \citep[UVOT; ][170--600 nm]{roming05}. The archival data from the Burst Alert Telescope \citep[BAT; ][15--150 keV]{barthelmy05}   from \citet{cusumano2010a} and \citet{cusumano2010b} were added to the SEDs as reference of the low  state.



\par The XRT data were reprocessed with standard procedures (\texttt{xrtpipeline v0.12.6}),
filtering and screening criteria by using the \texttt{Heasoft} package (v6.10).
We considered data collected using the photon counting (PC) mode with XRT event grades between 0 and 12.  Since the source count rate was always below 0.5 counts s$^{-1}$
no pile-up correction was necessary.
Source events were extracted from a circular region with a radius of 20
pixels (1 pixel $\sim$ 2.36$\arcsec$), while background events were extracted from a circular region with a radius of 50 pixels, away from background sources. Ancillary response files were generated
with \texttt{xrtmkarf}, and account for different extraction regions, vignetting and
PSF corrections. We used the spectral redistribution matrices v011 in the
calibration database maintained by HEASARC\footnote{http://heasarc.gsfc.nasa.gov/}.

All spectra were rebinned with a minimum of 20 counts per energy bin to allow
$\chi^2$ fitting within \texttt{XSPEC} (v12.6.0). We fit the individual
spectra with a simple absorbed power-law, with a neutral hydrogen column fixed
to its Galactic value (2.13$\times$10$^{20}$ cm$^{-2}$; Kalberla et
al. 2005). In addition, we summed the data collected after the $\gamma$-ray flare (2011 May 17-25) in order to have  better statistics see SED figures in Section \ref{sect:SED}.
The fit results are reported in Table \ref{swifttab} and the SED data points are reported in   Table \ref{tab:sedswift}.
During the 9 ToOs performed in 2011 April-May, \swift-XRT observed a 0.3--10 keV flux in the range (0.8-2.1) $\times$ 10$^{-11}$ erg
cm$^{-2}$ s$^{-1}$, a factor between 2 and 5 higher than the flux
level observed in 2008--2009. This is an hint that the mechanism that
produced an increase of the activity observed in $\gamma$-rays  also affected the X-ray part of the spectrum.

The peak of the X-ray flux was detected on 2011 May 15, soon after the major
$\gamma$-ray flare was detected by \fermi-LAT.
The flare  X-ray spectrum was softer than that of the post-flare  epoch,
thus demonstrating the contribution of the synchrotron component to lower energy X-ray emission.
This implies a shift of the synchrotron and IC peaks toward higher energies on  May 15.


\begin{figure}
 \centering
 \includegraphics[width=6cm,angle=-90]{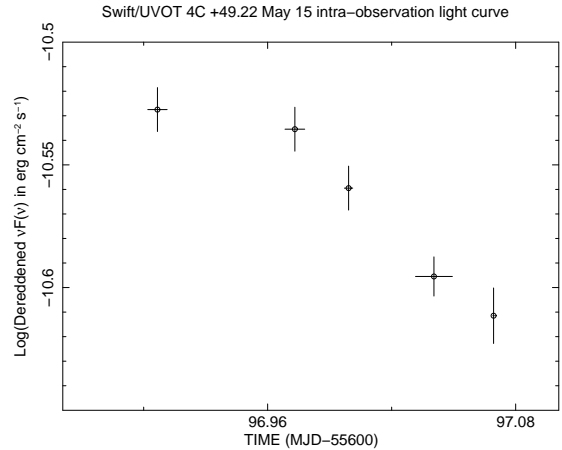}
 \caption{Photometric U-band  intra-day light curve obtained by \swift-UVOT in 2011 May 15.}
  \label{fig:uvot_variat15may}
\end{figure}

The \swift-UVOT can acquire images in six lenticular filters (V, B, U, UVW1, UVM2 and UVW2, with central wavelengths in the range 170--600 nm).
After seven years of operations, observations are now carried out using only one of the filters unless specifically requested by the user.
Therefore, images are not always available for all filters in all the observations.
The log of \swift-UVOT observations analysed is reported in Table \ref{swiftuvtab}.

The photometry analysis of all the \quattroc\ observations was performed using the standard UVOT
software distributed within the \texttt{HEAsoft} 6.9.0 package and the calibration included in the most recent release of the ``Calibration Database''.

We extracted source counts using a standard circular aperture with a 5\arcsec\ radius for all filters, and the background counts using an annular aperture with an inner radius of 26\arcsec\  and a width of 8\arcsec.
Source counts were converted to fluxes using the task \texttt{uvotsource} and the standard zero points \citep{poole08}.
Fluxes were then de-reddened using the appropriate values of $E(B-V)$ for the source taken
from \citet{schlegel98} and the $A_{\lambda}/E(B-V)$ ratios calculated for UVOT filters using the mean Galactic
interstellar extinction curve from \citet{fitzpatrick99}.


 Using U-band filter images only,
we detected variability  within a single exposure (a few hours timescale) in
the flare observation of 2011 May 15.
We show in Figure \ref{fig:uvot_variat15may} the U-band flux variation within this observation. If we compare the first ``segment'' of the exposure to the last, the variation is about 7$\sigma$.
This last segment is the shortest (34 s), and has the largest flux error.
%
%
Intra-day variability in a single  observation has
been detected also on May 19 in U,  and on May 22 in M2
filters. The May 19 observation includes three segments and shows
a flux increase on the last one (1230 s), while on May 22 the third
(191 s) of four segments shows a lower flux. The  May 19 episode
is the most significant, with a $\sim$10$\sigma$ variation.

\begin{table}
\caption[]{\swift-XRT spectral energy distribution data points}
\begin{center}
\label{tab:sedswift}
\begin{tabular}{lll}
\hline
Epoch          &  frequency   &  $\nu~f(\nu)$    \\

              &   [Hz]       & [erg cm$^{-2}$ s$^{-1}$]     \\
\hline
2011-05-15     &9.1$\times$10$^{16}$ &   (6.5$\pm$0.4)$\times$10$^{-12}$\\
(MJD: 55696)   &1.4$\times$10$^{17}$  &  (6.1$\pm$0.4)$\times$10$^{-12}$    \\
               &1.9$\times$10$^{17}$  &  (6.8$\pm$0.4)$\times$10$^{-12}$    \\
               &2.5$\times$10$^{17}$  &  (6.0$\pm$0.4)$\times$10$^{-12}$    \\
               &3.2$\times$10$^{17}$  &  (6.9$\pm$0.5)$\times$10$^{-12}$    \\
               &3.9$\times$10$^{17}$  &  (5.9$\pm$0.4)$\times$10$^{-12}$    \\
               &5.6$\times$10$^{17}$  &  (6.4$\pm$0.4)$\times$10$^{-12}$    \\
               &9.5$\times$10$^{17}$  &  (6.3$\pm$0.4)$\times$10$^{-12}$    \\
               &1.6$\times$10$^{18}$  & (6.5$\pm$1.5)$\times$10$^{-12}$\\
\hline
2011-05-17/25  & 8.8$\times$10$^{16}$& (2.1$\pm$0.1)$\times$10$^{-12}$\\
(MJD: 55698/55706)  & 1.3$\times$10$^{17}$& (2.2$\pm$0.1)$\times$10$^{-12}$\\
               & 1.7$\times$10$^{17}$& (2.4$\pm$0.1)$\times$10$^{-12}$\\
               & 2.1$\times$10$^{17}$& (2.5$\pm$0.1)$\times$10$^{-12}$\\
               & 2.5$\times$10$^{17}$& (2.8$\pm$0.2)$\times$10$^{-12}$\\
               & 2.9$\times$10$^{17}$& (2.7$\pm$0.1)$\times$10$^{-12}$\\
               & 3.4$\times$10$^{17}$& (2.8$\pm$0.1)$\times$10$^{-12}$\\
               & 4.1$\times$10$^{17}$& (3.0$\pm$0.2)$\times$10$^{-12}$\\
               & 4.9$\times$10$^{17}$& (3.1$\pm$0.2)$\times$10$^{-12}$\\
               & 6.1$\times$10$^{17}$& (3.3$\pm$0.2)$\times$10$^{-12}$\\
               & 7.8$\times$10$^{17}$& (3.7$\pm$0.2)$\times$10$^{-12}$\\
               & 1.0$\times$10$^{18}$& (3.9$\pm$0.2)$\times$10$^{-12}$\\
               & 1.5$\times$10$^{17}$& (4.1$\pm$0.3)$\times$10$^{-12}$\\
\hline
\end{tabular}
\end{center}
\end{table}


%
%
%
%
\section{Ground based and longer term radio-optical observations}\label{sect:radio-optical}
%
%

\subsection{MOJAVE  monitoring and component motion  studies}\label{sect:MOJAVE}
%

In order to study the parsec-scale morphology and possible changes in the
source structure, we analysed 13-epoch VLBA observations at 15 GHz from the
MOJAVE programme\footnote{The MOJAVE data archive is maintained at
http://www.physics.purdue.edu/MOJAVE.} spanning a time interval from
2008 May  to 2013 February. We imported the calibrated {\it uv}
data sets \citep{lister09} into the NRAO \texttt{AIPS}
package and
performed a few phase-only self-calibration iterations
before producing the final total intensity images. Uncertainties on
the flux density scale are within 5\%  \citep{lister13}. For the
six data sets obtained after the $\gamma$-ray flare we also produced
Stokes' Q and U images to study possible variations of the
source polarisation.  The uncertainties on the polarisation angle are less than 5$^{\circ}$ \citep{lister13}.\\

\begin{figure}
\centering
 \includegraphics[width=6cm]{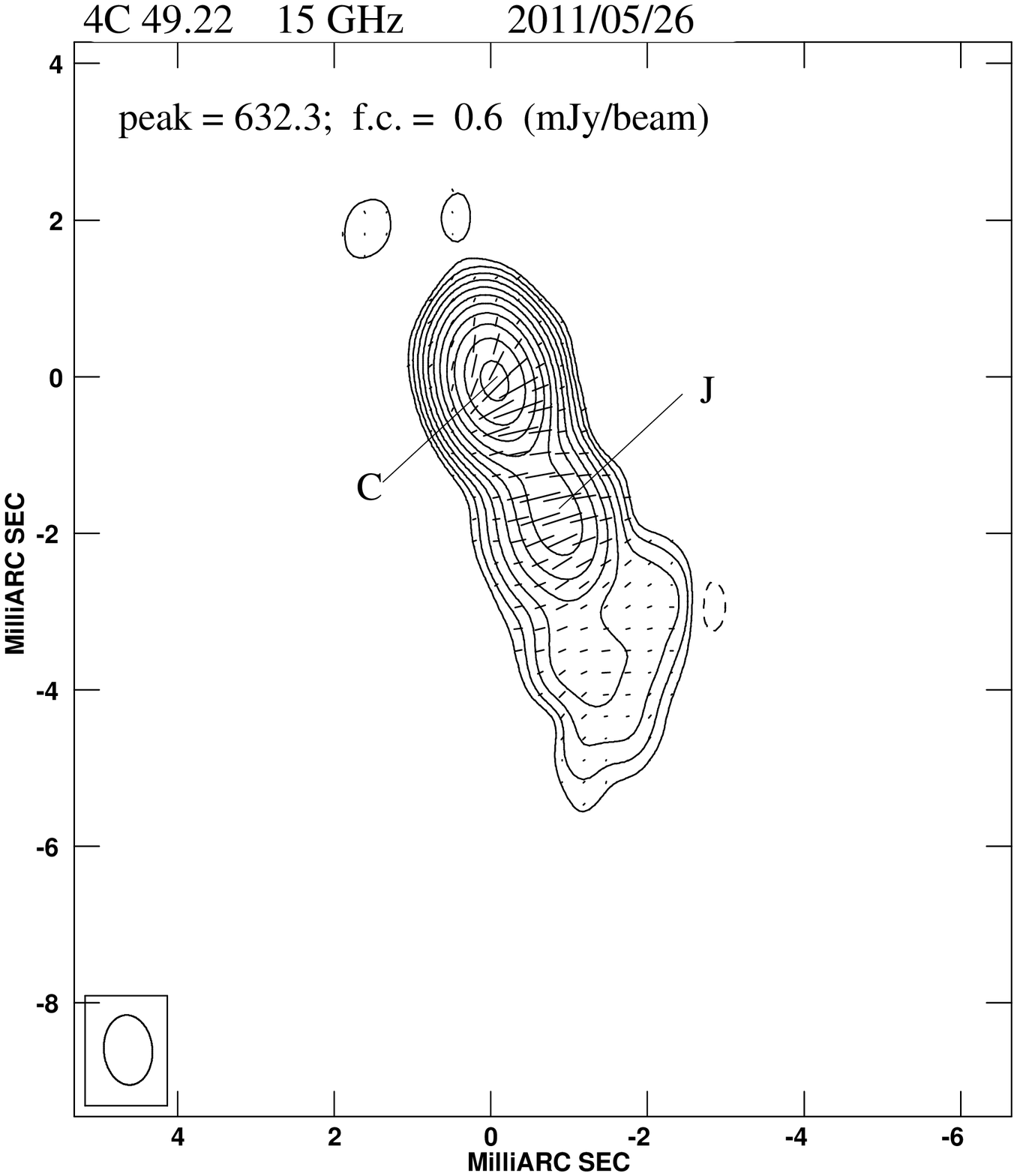}
 \includegraphics[width=6cm]{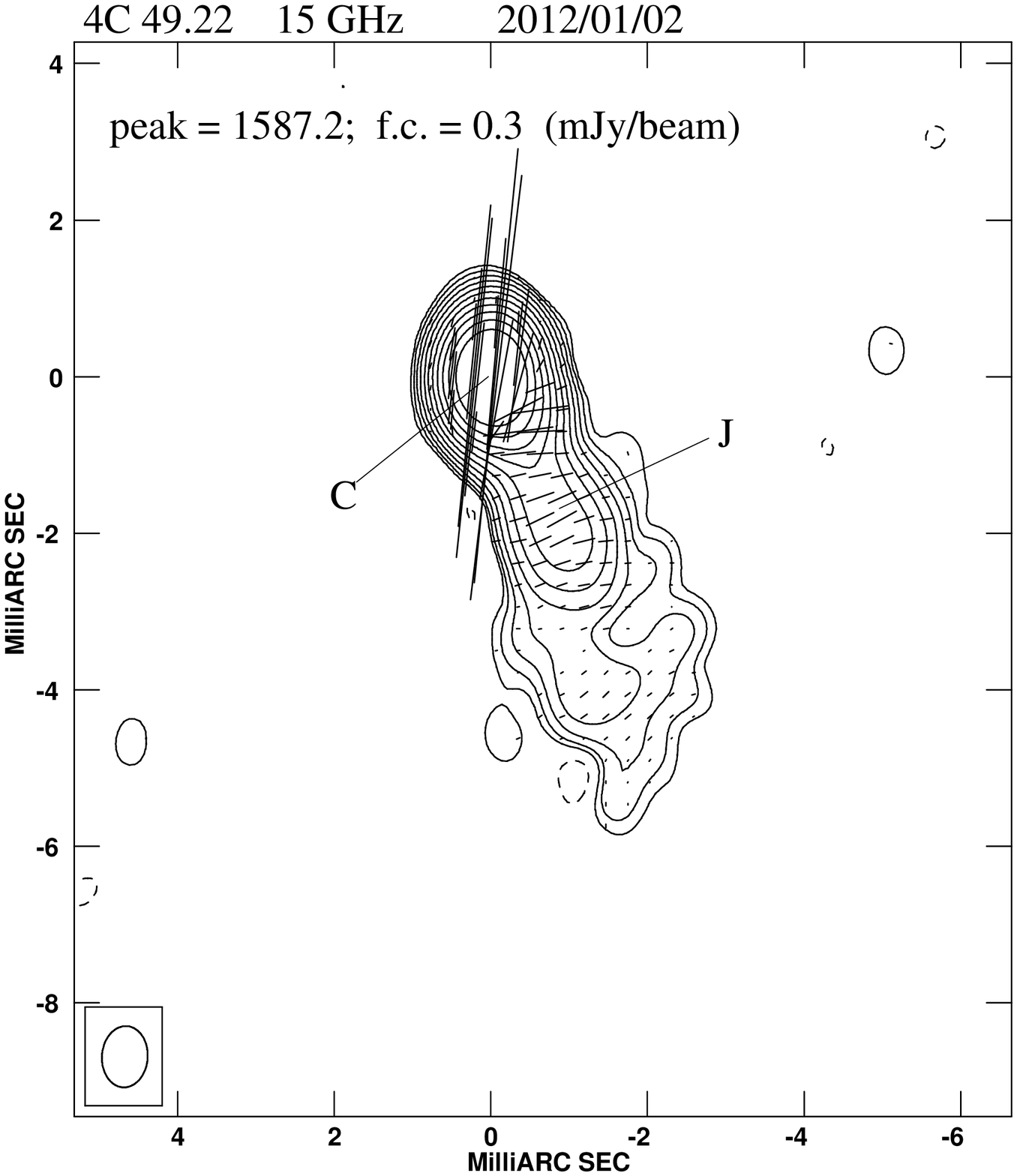}
\caption{15-GHz VLBA images of \quattroc\  from the observations in 2011 May
  (upper panel) and 2012 January (bottom panel).
On each image, we provide the observing date.
The peak flux density is in mJy/beam and the first contour (f.c.)
intensity is in mJy/beam, which corresponds to three times
the off-source noise level. Contour levels increase by a factor of 2.
The restoring beam is plotted in the bottom left-hand corner.
The vectors
superimposed on the total intensity flux density contours show the position angle of the electric
vector, where 1 mm length corresponds to 7.1 mJy/beam. }
\label{radio_structure}
\end{figure}

The source \quattroc\ shows a one-sided core-jet structure that is 6 mas in
size (i.e. $\sim$28 pc at the source redshift) and the radio emission is
dominated by the core component, labeled C in Figure \ref{radio_structure}.
Following a detection of the
$\gamma$-ray flare on 2011 May 15,
the core component of this source showed an increase in both the
total intensity of emission and the polarisation percentage, while the
polarisation angle has rotated by about 60$^{\circ}$. On the other
hand, no significant changes have been found in the jet structure, labeled J in Figure \ref{radio_structure},
strongly suggesting that the region
responsible for the radio variability is located within the central
component. Total intensity flux density and polarisation properties
of the core and jet components are reported in Table \ref{radio_flux}.

The total intensity and polarisation flux densities were
measured on the image plane with the Astronomical Image Processing System (AIPS\footnote{http://www.aips.nrao.edu}) using the Gaussian-profile fitting
task JMFIT and the task TVSTAT, which performs an aperture
integration on a selected region. As the source core we consider the
unresolved central component, and we derive its parameters with
JMFIT. The jet is the remaining structure, and the parameters are
obtained by subtracting the core contribution to the total emission
measured by TVSTAT. Errors are computed using the formulas from \citet{fanti01}.

To derive structural changes, in addition to the analysis performed on
the image plane, we also fitted the visibility data with  circular Gaussian components  at each epoch using the model-fitting option in
\texttt{DIFMAP}.
Errors $\Delta r$ associated with the component position are estimated by means of
$\Delta r = a/(S_{\rm p}/rms)$,  where $a$ is the component deconvolved major-axis,
$S_{\rm p}$ is its peak flux density and rms is the 1$\sigma$ noise level measured on the image plane
\citep{orienti2011}. In case the errors estimated are unreliably small,
 we assume a more conservative value for $\Delta r$ that is 10\% of the beam.

\begin{figure}
 \centering
\hspace{-1cm}
 \includegraphics[width=6cm]{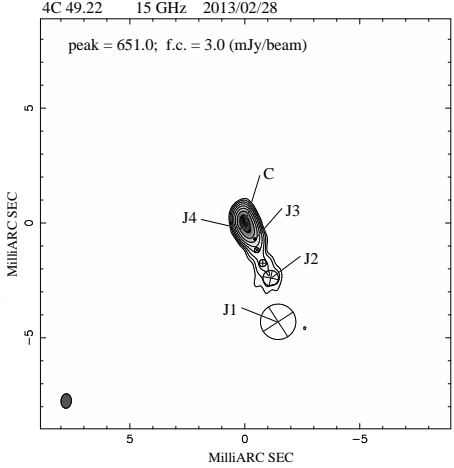}
\caption{
15-GHz VLBA image from the
observations performed during 2013 February with the positions of the  of the jet components discussed in Section 5.1.
The peak flux density is in mJy/beam and the first contour (f.c.) intensity is in mJy/beam, and it corresponds to 0.5 \% of the peak flux density.
Contour levels increase by a factor of 2. The restoring beam is plotted in the bottom left-hand corner.}
\label{modelfit}
\end{figure}

\begin{figure}
 \centering
\hspace{-1cm}
 \includegraphics[width=6cm]{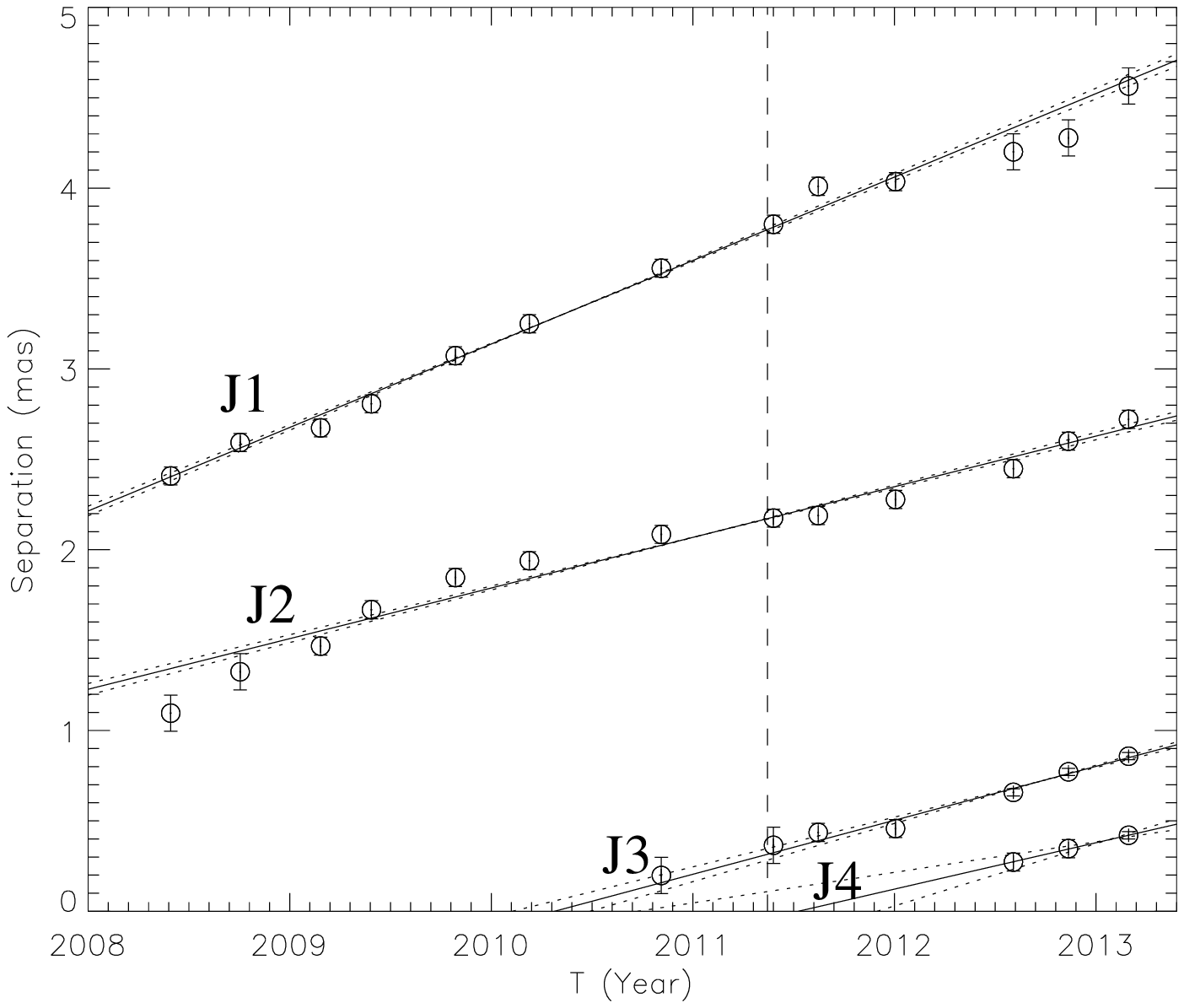}
\caption{
Changes in separation with time between
component C, considered stationary, and J1, J2, J3 and J4. The solid lines represent the
linear fits to the data, while the dotted lines represent the
uncertainties from the fit parameters. The vertical dashed line
indicates the time of the $\gamma$-ray flare.}
\label{motion}
\end{figure}

This approach is preferable in order  to derive
small variations in the source structure; it also provides a more accurate
fit of unresolved structures close to the core component.
Throughout the observing epochs we could reliably follow the motion
of only two components, labeled J1 and J2 in Figure \ref{modelfit};
a third component, J3, became visible in the last seven
epochs. Interestingly, a new component, J4, was detected in the last
three epochs of MOJAVE data, since 2012 August. We determined the separation velocity from the core, considered stationary, for  these four
components by means of a linear regression fit that minimises the
chi-square error statistics (Figure \ref{motion}).
From this analysis we found that J1, J2, J3 and J4 are increasing their separation from the core with  apparent
angular velocities of 0.48$\pm$0.01, 0.27$\pm$0.01,  0.30$\pm$0.02  and 0.27$\pm$0.09 mas/yr, which correspond to  apparent linear velocities
$\beta_{\rm app}=v_{app}/c$ of 9.9$\pm$0.2, 5.6$\pm$0.2, 6.2$\pm$0.4, and
5.6$\pm$1.4, respectively.
The velocity derived for J1 is in agreement with the value found by \citet{lister13}, while for J2 we obtain a slower speed. This may be due to a deceleration that may become detectable with the availability of additional observing epochs not considered in \citet{lister13}.
The large uncertainty on the velocity of
component J4 is due to the availability of only 3 observing epochs
spanning a very short time interval of about 7 months.
From a linear regression fit we estimated the time of zero separation, which provides an indication of the ejection epoch. We found that J4 emerged
from the core on 2011.52 (i.e., beginning of July 2011), making the
ejection of the component close in time to the $\gamma$-ray flare.
However, the large uncertainties on the separation velocity
do not allow us to accurately constrain the precise time of zero
separation, which ranges between 2010.73 (i.e. 2010 September) and
2011.92 (i.e. 2011 December). It is worth noting that the
time of zero separation estimated for component J3 is
2010.3$\pm$0.2 (i.e. 2010 April) with an associated uncertainty on
the ejection time that
ranges between 2010 February and June. Interestingly, no strong
$\gamma$-ray flare was reported close in time with the ejection of
this component, but the source turned out to be repeatedly detectable by LAT on a weekly time scale after 2010 February  (see Section \ref{sect:gamma-lightcurve}).
By means of the apparent velocities derived for the jet components, we
estimated the possible combination of the intrinsic velocity $\beta=v/c$
and the angle $\theta$ that the jet forms with our line of sight:

\begin{equation}
\beta_{\rm app} = \frac{\beta {\rm sin} \theta}{1 - \beta {\rm
cos} \theta}
\label{velocity}
\end{equation}
%
\noindent    We assumed that $\beta_{\rm app}$ is between 9.9 and 5.6, i.e. the velocity estimated for the fastest and slowest components.
In the former case we found that $\beta >0.995$ and $\theta <11^{\circ}$, while in the latter $\beta > 0.985$ and $\theta < 20^{\circ}$.
Another way to derive the possible ($\beta$-$\theta$)
combinations is using the flux density ratio of the approaching,
$S_{\rm a}$, and
receding $S_{\rm r}$ jets:

\begin{equation}
\frac{S_{\rm a}}{S_{\rm r}}= \left( \frac{1+ \beta {\rm cos} \theta}{1
- \beta {\rm cos} \theta} \right)^{3+ \alpha}
\label{ratio}
\end{equation}

\noindent   where $\alpha$ is the spectral index ($S_{\nu} \propto \nu^{- \alpha}$), and it is assumed to be 0.7, i.e. a typical value for the jet component. From the lack of detection of a counterjet in any of the images, we set a lower limit on the jet/counterjet ratio. We used the highest ratio between the jet and counterjet emission,  which is computed by using $S_{\rm a} = 816$ mJy (i.e. the flux density of J4 as it emerges from the core, see Table \ref{radio_flux}), and $S_{\rm r} = 0.3$ mJy, which corresponds to 1$\sigma$ rms. From these values we derive a jet/counterjet ratio $>$2720. This value yields $\beta$cos$\theta >$0.79 c, implying $\beta >0.79$ and $\theta < 38^{\circ}$,  which are consistent with the range found from Equation \ref{velocity}.


With the derived values we can compute a  lower limit on the Doppler factor by means of:

\begin{equation}
\mathcal{D} = \frac{1}{\Gamma (1 - \beta {\rm cos} \theta)}
\label{doppler}
\end{equation}

\noindent
where $\Gamma$ is the bulk Lorentz factor. The lower limit on the Doppler factor is $\mathcal{D} > 4.2$, which is compatible with those derived from our SED modeling (see Section \ref{sect:SED}). However, this estimate is affected by the large uncertainties in the apparent velocity and more observations spanning a larger time interval with  frequent time sampling are necessary. Our SED modeling results (Section \ref{sect:SED}) are constrained to $\mathcal{D}$ values between $\sim$20 and $\sim$30, maintaining therefore the agreement with variability Doppler factors obtained recently for a sample of \fermi-LAT blazars \citep[e.g., ][]{pushkarev09,savolainen10}, and in particular  maintaining the agreement with previous SED modeling of \quattroc\ \citep{sambruna06b}.
The observed increasing flux density and polarisation degree in the
radio core of \quattroc\ after the GeV $\gamma$-ray flare demonstrate
that the high-energy peak emission is produced in or close to the radio core
rather than in structures and blobs at larger distance along the jet, far from the central engine.


Even if this does not yet constrain the flaring GeV emission
region (sometimes called  the ``blazar zone'') to within the BLR, we can at least exclude flaring GeV emission produced by jet knots placed at a large
distance from the central engine \cite[e.g., few parsecs][]{lister13}.
This result can help to discriminate
between our multi-temporal and multi-wavelength SED
modeling described in Section \ref{sect:SED}.


\begin{table*}
\caption{Log of the data obtained by
the UVOT instrument on board \swift. All magnitudes are corrected for Galactic extinction.}
\begin{center}
\begin{tabular}{cccccccc}
\hline
Date& Exp. Time &    V  & B & U  & W1 & M2 & W2 \\
    & (s)       &  (Mag) & (Mag) & (Mag)      & (Mag)   & (Mag)    & (Mag) \\
\hline
2008-04-28 (MJD: 54584) & 5398  &  --  & -- & -- & 15.64$\pm$0.04 & --  & -- \\
2009-05-06 (MJD: 54957) & 2980  &  17.62$\pm$0.09  & 17.59$\pm$0.05 & 16.40$\pm$0.04 & 16.42$\pm$0.05 & 15.91$\pm$0.05  & 16.06$\pm$0.04 \\
2009-11-17 (MJD: 55152) & 5034  &  17.34$\pm$0.07  & 17.18$\pm$0.04 & 16.01$\pm$0.03 & 15.72$\pm$0.04 & 15.28$\pm$0.04  & 15.41$\pm$0.03 \\
2011-04-26  (MJD: 55677) & 1336  &  --  & -- & -- & 14.77$\pm$0.04 & --  & -- \\
2011-04-29  (MJD: 55680) & 4750  &  16.37$\pm$0.04  & 16.55$\pm$0.04 & 15.55$\pm$0.04 & 15.42$\pm$0.04 & 15.10$\pm$0.04  & 15.22$\pm$0.04 \\
2011-05-02  (MJD: 55683) & 4247  &  15.97$\pm$0.04  & 16.27$\pm$0.04 & 15.21$\pm$0.04 & 15.12$\pm$0.04 & 14.82$\pm$0.04  & 14.96$\pm$0.04 \\
2011-05-15  (MJD: 55696) & 3592  &  --  & -- & 14.14$\pm$0.02 & -- & --  & -- \\
2011-05-17  (MJD: 55698) & 3390  &  15.50$\pm$0.03  & 15.70$\pm$0.02 & 14.79$\pm$0.02 & 14.73$\pm$0.03 & 14.56$\pm$0.04  & 14.70$\pm$0.03 \\
2011-05-19  (MJD: 55700) & 3689  &  --  & -- & 14.25$\pm$0.02 & -- & --  & -- \\
2011-05-22  (MJD: 55703) & 4077  &  --  & -- & -- & 15.14$\pm$0.03 & 14.87$\pm$0.03  & 15.04$\pm$0.03 \\
2011-05-23  (MJD: 55704) & 3936  &  --  & -- & -- & 14.70$\pm$0.03 & 14.47$\pm$0.03  & 14.63$\pm$0.03 \\
2011-05-25  (MJD: 55706) & 3849  &  --  & -- & -- & 15.28$\pm$0.03 & 14.97$\pm$0.03  & 15.15$\pm$0.03 \\
\hline
\end{tabular}
\end{center}
\label{swiftuvtab}
\end{table*}

\begin{table*}
\caption{Total intensity flux density and polarisation
properties. Column 1: observing date; Cols. 2, and 3: total intensity
flux density of component C and J, respectively; Cols. 4, and 5:
polarised flux density (and polarisation percentage) for component C
and J, respectively; Cols. 6, and 7: polarisation angle for component
C and J, respectively.}
\begin{center}
\begin{tabular}{ccccccc}
\hline
Date&$S_{\rm C}$&$S_{\rm J}$&$S_{\rm pol,C}$&$S_{\rm
pol,J}$&$\chi_{\rm C}$&$\chi_{\rm J}$\\
 & mJy & mJy& mJy (\%) & mJy (\%) & deg & deg \\
\hline
&&&&&&\\
2011-05-26 (MJD: 55707)& 658$\pm$33& 88$\pm$5& 5.8$\pm$0.5 (0.9$\pm$0.1\%)& 7.2$\pm$0.7 (8.1$\pm$0.8\%)& -64$\pm$5& -66$\pm$5\\
2011-08-15 (MJD: 55788)& 947$\pm$47& 91$\pm$5& 9.4$\pm$0.6 (1.0$\pm$0.1\%)& 7.4$\pm$0.7 (8.1$\pm$0.8\%)& -84$\pm$5& -77$\pm$5\\
2012-01-02 (MJD: 55928)&1708$\pm$85& 87$\pm$4&45.1$\pm$2.2 (2.6$\pm$0.1\%)& 8.2$\pm$0.8 (9.4$\pm$0.8\%)& -10$\pm$5& -73$\pm$5\\
2012-08-03 (MJD: 56142)&1650$\pm$82& 92$\pm$5&59.0$\pm$3.0 (3.5$\pm$0.2\%)& 6.0$\pm$0.7 (6.5$\pm$0.7\%)& 10$\pm$5& -62$\pm$5\\
2012-11-11 (MJD: 56242)&1381$\pm$69& 63$\pm$3&62.0$\pm$3.1 (4.5$\pm$0.2\%)& 5.0$\pm$0.6 (7.9$\pm$0.9\%)& 19$\pm$5& -67$\pm$5\\
2013-02-28 (MJD: 56351)&1081$\pm$54& 65$\pm$3&20.0$\pm$1.1 (1.9$\pm$0.1\%)& 6.0$\pm$0.7 (9.2$\pm$1.0\%)& 48$\pm$5& -70$\pm$5\\

&&&&&&\\
\hline
\end{tabular}
\end{center}
\label{radio_flux}
\end{table*}

\subsection{Mets\"{a}hovi Radio Observatory}\label{sect:mets}

Observations at 37 GHz were made with the 13.7 m diameter Mets\"ahovi
radio telescope, which is a radome-enclosed paraboloid antenna situated in
Finland. The measurements were made with a 1 GHz-band dual beam receiver
centred at 36.8 GHz. The observations are ON--ON observations,
alternating the source and the sky in each feed horn. A typical
integration time to obtain one flux density data point is between 1200 and
1400 s. The detection limit of the telescope at 37 GHz is on the order of
0.2 Jy under optimal conditions. Data points with a signal-to-noise ratio
$<$ 4 are handled as non-detections. The flux density scale is set by
observations
of DR 21. Sources NGC 7027, M 87 (3C 274) and NGC 1275 (Per A, 3C 84) are used as secondary
calibrators. A detailed description of the data reduction and analysis is
given in \citet{terasranta98}.  The error estimate in the flux density includes
the contribution from the measurement rms and the uncertainty of the
absolute calibration.
The  flux density light curve is shown in Figure \ref{wholelcLAT} and Table \ref{tab:mets}.

\begin{table}
\caption[]{Flux densities of \quattroc\ with Mets\"{a}hovi Radio Observator at 37 GHz} 
\begin{tabular}{cc}
\hline
Date &    Flux Density     \\

     &      [mJy]     \\
\hline

2008-08-06 (MJD: 54684.0)   & 0.90$\pm$0.08\\
2009-01-03 (MJD: 54834.0)   & 1.12$\pm$0.09\\
2009-02-10 (MJD: 54872.0)   & 0.90$\pm$0.12\\
2009-04-23 (MJD: 54944.0)   & 1.10$\pm$0.09\\
2009-09-20 (MJD: 55094.0)   & 0.63$\pm$0.11\\
2009-11-12 (MJD: 55147.0)   & 0.90$\pm$0.13\\
2009-11-30 (MJD: 55165.0)   & 1.25$\pm$0.11\\
2009-12-12 (MJD: 55177.0)   & 0.81$\pm$0.11\\
2010-02-25 (MJD: 55252.0)   & 1.54$\pm$0.1 \\
2010-05-12 (MJD: 55328.0)   & 1.41$\pm$0.13\\
2010-05-25 (MJD: 55341.0)   & 1.27$\pm$0.09\\
2010-06-27 (MJD: 55374.0)   & 1.92$\pm$0.11\\
2010-11-07 (MJD: 55507.0)   & 1.94$\pm$0.3 \\
2011-02-03 (MJD: 55595.0)   & 1.14$\pm$0.1 \\
2011-02-06 (MJD: 55598.0)   & 1.03$\pm$0.14\\
2011-03-18 (MJD: 55638.0)   & 1.26$\pm$0.11\\
2011-05-18 (MJD: 55699.0)   & 1.45$\pm$0.14\\
2011-05-22 (MJD: 55703.0)   & 1.71$\pm$0.09\\
2011-05-26 (MJD: 55707.0)   & 1.57$\pm$0.17\\
2011-05-27 (MJD: 55708.0)   & 1.46$\pm$0.1 \\
2011-05-28 (MJD: 55709.0)   & 1.61$\pm$0.19\\
2011-05-29 (MJD: 55710.0)   & 1.42$\pm$0.08\\
2011-05-30 (MJD: 55711.0)   & 1.52$\pm$0.14\\
2011-06-01 (MJD: 55713.0)   & 1.22$\pm$0.09\\
2011-06-02 (MJD: 55714.0)   & 1.61$\pm$0.09\\
2011-06-05 (MJD: 55717.0)   & 1.53$\pm$0.09\\
2011-07-24 (MJD: 55766.0)   & 1.91$\pm$0.28\\
2011-08-10 (MJD: 55783.0)   & 2.07$\pm$0.09\\

\hline

\label{tab:mets}
\end{tabular}
\end{table}

\subsection{Quasi-Simultaneous near-infrared monitoring by INAOE}\label{sect:INAOE}
%
%

The Near-infrared (NIR) photometry was performed in J, $K_s$ and H band with the CANICA
NIR camera at the 2.1-m telescope of the Observatorio
Astrof\'{\i}sico Guillermo Haro (OAGH) in Cananea, Sonora,
Mexico. CANICA is a camera based on a HAWAII 1024 $\times$ 1024 pixel
array, with plate scale and field of view  0.32 \arcsec\ pixel$^{-1}$
and about 5.5 $\times$ 5.5 square arcmin, respectively. Observations were
reduced using standard differential aperture photometry with IRAF
packages\footnote{http://iraf.noao.edu/}. Every night, several standard stars from \citet{hunt98} were
observed in both bands. The photometric error for each night was
assumed to be the standard deviation between our estimated magnitude
and the magnitude determined in \citet{hunt98} for the standard stars
observed. The mean zero-point error in J and H in the photometric nights is
0.06 mag, reaching 0.08 mag on the photometric nights. Similarly,
in $K_s$ we have errors of 0.08 and 0.10 mag, respectively. As expected,
the error estimates show that the photometric accuracy is generally
higher in J and H than in $K_s$.  We note that our average errors in the
photometric zero-point are 0.07 mag in J and H and 0.09 mag in $K_s$, which is
quite good for NIR bands. The flux densities in the three filters are reported 
in Figure \ref{fig:LC} and in Table \ref{tab:inaoe}.




Each image was checked for problems before going through all the data-reduction steps. For the
treatment of the images, we used the GEMINI IRAF1 package. Flat-field
images were obtained through the QFLAT task from combining many
dome images. We decided to use dome flat-fields instead of flat
images obtained from sky images; after checking with many standard
stars, we discovered that the former produced more accurate results in terms of
photometry (i.e. smaller zero-point errors). The QSKY task was used
to estimate the background contribution. For each image,
the background was estimated from the four sky images closest in
time. Furthermore, the mean and standard deviation of each sky
image were calculated, and if the mean was discrepant from that
of the other three sky images by more than 10 \% of the standard
deviation then the image was removed from the process. In this
way, we avoid background changes and ensure a proper background
subtraction. These corrections (flat-fielding and background
subtraction) were performed for every galaxy and standard star image
by QREDUCE, using the appropriate flat-field and background
images. Finally, the IMCOADD task combines all corrected
images of a galaxy according to their median, calculating the necessary shifts due to the
dither pattern.

\begin{table}
\caption[]{Magnitudes and flux densities of \quattroc\ with INAOE Telescope} 
\begin{tabular}{cccc}
\hline
Filters& Date   & Magnitude & flux    \\

     &           &          & [mJy]     \\
\hline

	J  &2011-04-29 (MJD: 55680.84) &	14.45$\pm$0.07 & 		2.66$\pm$0.32	  \\
	  &2011-05-09 (MJD: 55690.78) &	14.55$\pm$0.07 & 		2.41$\pm$0.29	  \\
	  &2011-05-10 (MJD: 55691.77) &	14.29$\pm$0.07 & 		3.06$\pm$0.37	  \\
	  &2011-05-14 (MJD: 55695.75) &	13.33$\pm$0.07 & 		7.46$\pm$0.92	  \\
	  &2011-05-21 (MJD: 55702.77) &	13.42$\pm$0.07 & 		6.83$\pm$0.84	  \\
	  &2011-05-23 (MJD: 55704.69) &	13.63$\pm$0.09 & 		5.62$\pm$0.89	  \\
	  &2011-05-24 (MJD: 55705.74) &	13.68$\pm$0.05 & 		5.40$\pm$0.47	  \\
	H  &2011-04-29 (MJD: 55680.82) &	13.50$\pm$0.08 & 		4.26$\pm$0.90	  \\
	  &2011-05-09 (MJD: 55690.78) &	13.80$\pm$0.09 & 		3.24$\pm$0.77	  \\
	  &2011-05-10 (MJD: 55691.76) &	13.42$\pm$0.08 & 		4.56$\pm$0.96	  \\
	  &2011-05-13 (MJD: 55694.77) &	13.04$\pm$0.08 & 		6.51$\pm$1.37	  \\
	  &2011-05-14 (MJD: 55695.75) &	12.49$\pm$0.07 & 		10.8$\pm$1.99  \\
	  &2011-05-21 (MJD: 55702.76) &	12.40$\pm$0.07 & 		11.7$\pm$2.16  \\
	  &2011-05-23 (MJD: 55704.69) &	12.76$\pm$0.06 & 		8.44$\pm$1.33	  \\
	  &2011-05-24 (MJD: 55705.74) &	12.89$\pm$0.06 & 		7.46$\pm$1.18	  \\
	Ks &2011-04-29 (MJD: 55680.85) &	12.50$\pm$0.11 & 		6.61$\pm$3.09	  \\
	 &2011-05-09 (MJD: 55690.79) &	12.73$\pm$0.06 & 		5.37$\pm$1.37	  \\
	 &2011-05-10 (MJD: 55691.78) &	12.62$\pm$0.12 & 		5.95$\pm$3.03	  \\
	 &2011-05-14 (MJD: 55695.75) &	11.53$\pm$0.12 & 		16.2$\pm$8.30  \\
	 &2011-05-21 (MJD: 55702.78) &	11.86$\pm$0.12 & 		12.0$\pm$6.12  \\
	 &2011-05-23 (MJD: 55704.69) &	12.07$\pm$0.12 & 		9.91$\pm$5.05	  \\
	 &2011-05-24 (MJD: 55705.74) &	12.13$\pm$0.09 & 		9.36$\pm$3.58	  \\

\hline

\label{tab:inaoe}
\end{tabular}
\end{table}

\subsection{Catalina Sky Survey observations}\label{sect:catalina}
%

The Catalina Sky Survey  for near-Earth objects and potential planetary
 hazard asteroids (NEO/PHA) is conducted by the University of
Arizona Lunar and Planetary Laboratory group\footnote{S. Larson, E. Beshore, and collaborators; see
http://www.lpl.arizona.edu/css/}. CSS utilizes three wide-field
telescopes: the 0.68-m Catalina Schmidt at Catalina Station, AZ; the
0.5-m Uppsala Schmidt (Siding Spring Survey, or SSS, in collaboration
with the Australian National University) at Siding Spring Observatory,
NSW, Australia; and the Mt. Lemmon Survey (MLS), a 1.5-m reflector
located on Mt. Lemmon, AZ. Each telescope employs a camera with a
single, cooled, 4k$\times$4k back-illuminated, unfiltered CCD.
Between the three telescopes, the majority of the observable sky is
covered at least once (and up to 4 times) per lunation, depending on
the time since the area was last surveyed and its proximity to the
ecliptic. The total area coverage is   $\sim$ 30,000 deg$^2$, and it excludes
the Galactic plane within $|b| < 10^{\circ}$. For each coverage four images of the same
field are taken, separated in time by $\sim$ 10 min, for a total time
baseline of $\sim$ 30 min in that sequence. Typically 2 to 4 such
sequences are obtained per field per lunation; the cycle is generally
repeated the next lunation, marching through the RA range during the
year. The time baselines now extend to 6 years with up to $\sim$ 300
exposures per pointing over much of the area surveyed so far. This
represents an unprecedented coverage in terms of the combined area,
depth, and number of epochs. The photometric flux data of
\quattroc\ are retrieved through the Catalina Surveys Data Release
services\footnote{http://nesssi.cacr.caltech.edu/DataRelease/}  (see Figure \ref{wholelcLAT}).

%
%
%
\section{$\gamma$-ray flare and time variability}\label{sect:gamma-lightcurve}
%
%

\begin{figure*}
 \centering
 \includegraphics[width=\hsize,angle=0]{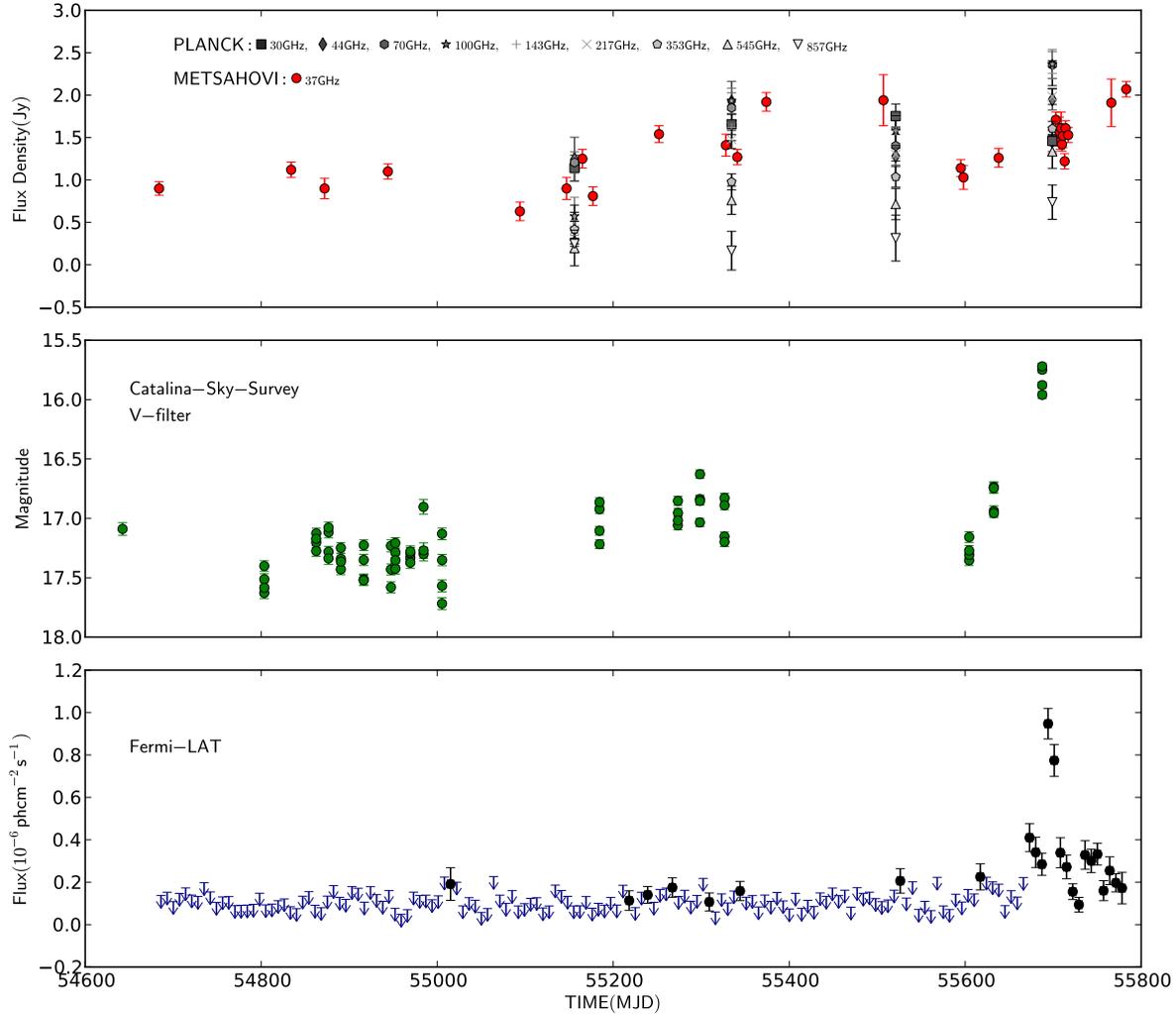}
 \caption{Multi-frequency light curves using the long term monitoring data.
Top panel: Mets\"{a}hovi radio data  and \planck\ (described in Sect. \ref{sect:mets} and in Sect. \ref{sect:planck}), middle panel: Catalina Sky Survey (described in Sect. \ref{sect:catalina}), bottom panel:
36 months $\gamma$-ray integrated flux ($E > 100$ MeV) lightcurve  measured by \fermi-LAT from 2008 August 4 to 2011 August 4.
The time binning is 7 days: blue arrows represent the 2-$\sigma$ upper limits (described in Sect. \ref{sect:LAT}).}
  \label{wholelcLAT}
\end{figure*}
\begin{figure*}
 \centering
\hspace{-1cm}
 \includegraphics[width=\hsize,angle=0]{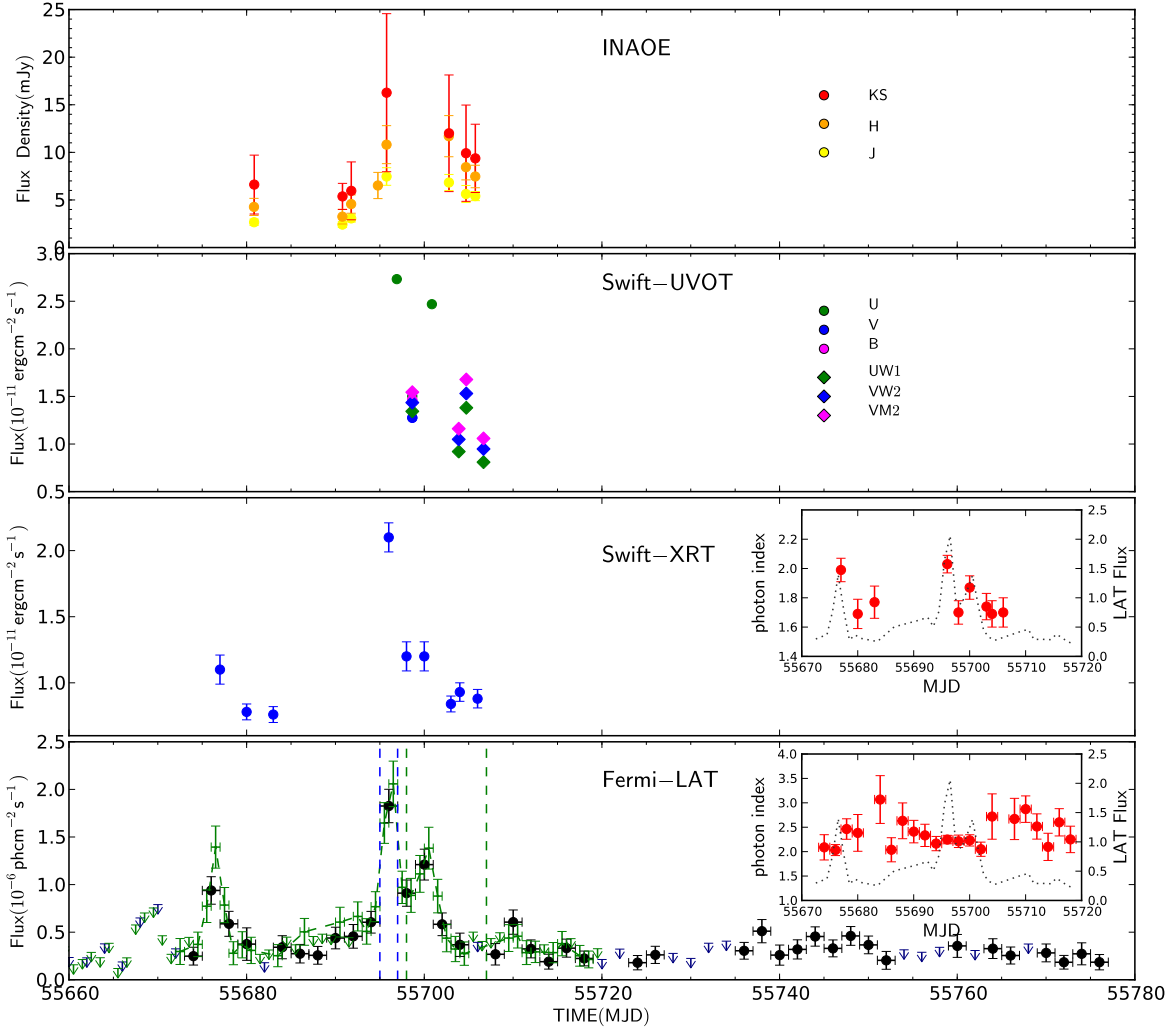}
 \caption{Multi-frequency light curves  zoomed in to the main flaring period (2011 May). In the first panel INAOE data points are shown in the different filters (described in Sect. \ref{sect:INAOE}); in the second panel, \swift-UVOT (described in Sect. \ref{sect:swift}); in the third panel, \swift-XRT (described in Sect. \ref{sect:swift}); and in the fourth panel, \fermi-LAT 1 day time bin data, are shown (described in Sect. \ref{sect:LAT}). The blue and green dashed lines represent the two states considered in the SED modeling. The inset panels report the spectral indices coming from the power-law spectral fit in each bin with superimposed the \fermi-LAT $\gamma$-ray lightcurve.}
  \label{fig:LC}
\end{figure*}

To investigate the
behaviour of this source, in particular the flaring state phase, we
extracted the light curves from the entire  data set using different time
binnings (1 week, 3 days and 1 day time bins) with \texttt{gtlike}. The source spectrum was fitted with a
power-law  function. In the case of 1 day time bins, we fixed the photon index to the value found in the whole energy band, integrating over 3 years of data.
For longer time bins the photon index was left free to vary.
The lowest panel of Figure  \ref{wholelcLAT}  shows the whole \fermi-LAT lightcurve with a weekly time bins. We report in the same figure radio and optical long-term observations by Mets\"{a}hovi and \planck\ and Catalina Sky Survey.
Figure \ref{fig:LC} shows the  parts of the multi-frequency light curves during the main flaring activity, increasing in frequency from the top to the bottom panel.
We studied the evolution of spectral shape during the flaring state for the X-ray and  $\gamma$-ray components. We show in Figure \ref{fig:LC} the photon index values versus time with overlayed \fermi-LAT lightcurve shape in the inset panels. We performed a linear regression to evaluate the dependence between flux values and spectral indices, for \fermi-LAT no spectral evolution is evident during the flaring state with a regression coeffient of $r\sim0.01$, but for the X-rays, a softer when brighter behaviour consistent with a shift in the synchrotron peak during the flaring state is  noticeable ($r\sim0.74$).

\begin{figure}
 \centering
 \includegraphics[width=6cm,angle=270]{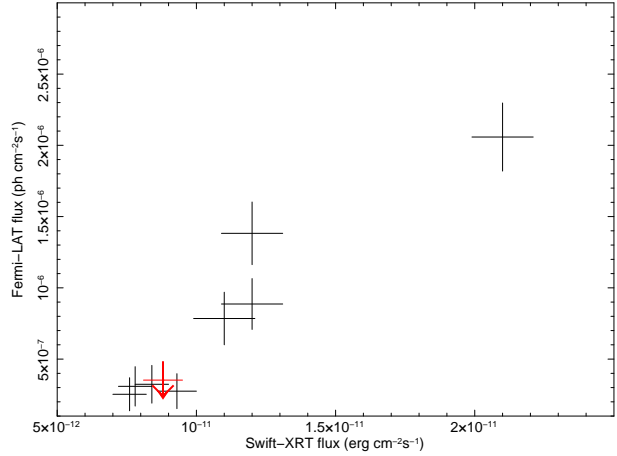}
 \caption{$\gamma$-ray vs. X-ray flux at the times of \swift-XRT observations.
\fermi-LAT fluxes were interpolated from the 1 day binned $\gamma$-ray light curve.}
  \label{gamma_x_ray}
\end{figure}

\begin{figure}
 \centering
 \includegraphics[width=7cm,angle=0]{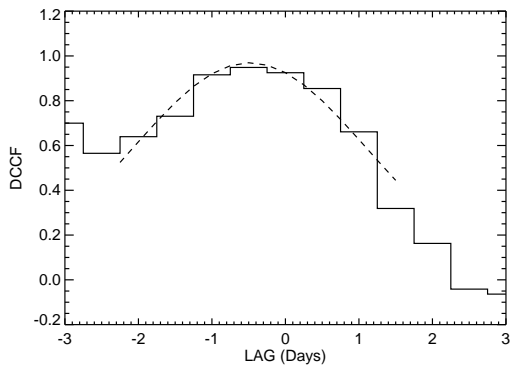}
 \caption{Discrete Cross Correlation Function using the \swift-XRT data and the \fermi-LAT 1 day binned $\gamma$-ray lightcurve, The dashed  curve is a Gaussian fit to the DCCF peak. The time lag is estimated to be $-0.4 \pm 1.0$ days
(negative means X-rays leading $\gamma$-rays).}
  \label{DCCF}
\end{figure}

A direct comparison of the \fermi-LAT and CSS light curves clearly shows that an optical brightening occurred at the time of the $\gamma$-ray flare.
Since no optical data for the same filter are available for the post-flare relaxing phase we cannot estimate any time lag.
Although the \swift-XRT data only cover a limited time range  (see Figure \ref{fig:LC}) these
observations suggest a strong correlation between the X-ray and
GeV flux. This correlation is illustrated in Figure \ref{gamma_x_ray} where we have
plotted the $\gamma$-ray versus X-ray flux at the times of the X-ray
\swift-XRT measurements.
During the  period of the last  \swift-XRT observation the source was not detected by \fermi-LAT  and we did also include the upper limit in Figure \ref{gamma_x_ray}.
The $\gamma$-ray flux for each point was obtained by
linear interpolation of the 1-day bin LAT light curve.
With only 9 X-ray points available a standard discrete cross
correlation function (DCCF) would be very poorly sampled.
On timescales longer than the flare lengths of a few days the
DCCF contains no significant information. It is however possible to use the
DCCF on  shorter timescales to estimate a time lag between the X-ray
and $\gamma$-ray variations. In Figure \ref{DCCF} we show the DCCF for lags less
than 3 days computed by oversampling the 1 day binned LAT light curve by
a  factor of 19,  to give a larger oversampling, before correlating it with the X-ray points.
The Gaussian fit, which is also shown in the figure, gives an estimated
time lag $\Delta t = -0.4 \pm 1.0$ days (where negative lag means X-rays
preceding $\gamma$-rays). Uncertainty estimates were made by two
different Monte Carlo methods: the model independent approach
by \citet{peterson98} and by simulated flare light curves.
In the second approach double sided exponential flares were sampled in a similar manner
to the observations in order to investigate uncertainties due
to the precise timing of the X-ray observations. The two methods gave consistent estimates.



\begin{figure}
 \centering
 \includegraphics[width=6cm,angle=270]{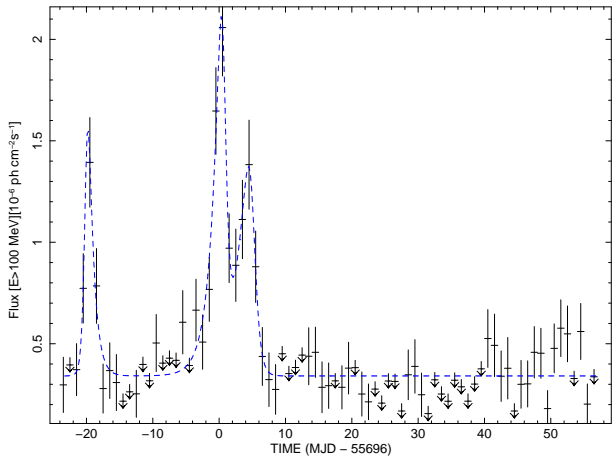}
 \caption{One day time bin light curve of \fermi-LAT data zoomed in to the flaring period of 2011 May  with superimposed  fitting function shown as a blue dashed line.}
  \label{fit_lc}
\end{figure}

We analysed the $\gamma$-ray flare of 2011 May 15 (MJD 55696), using the following function $F(t)$, already proposed in \citet{variability}, to fit the $\gamma$-ray light curve shape during each single flare:

\begin{equation}
F(t)=F_c+F_0\left(e^{\frac{t_0-t}{T_r}}+e^{\frac{t-t_0}{T_d}}\right)^{-1}
\end{equation}
%
Here, $F_c$ represents the baseline of the flux lightcurve, $F_0$ measures the amplitude of
the flares, $t_0$ describes approximatively the time
of the peak (it corresponds to the actual maximum only for symmetric flares), $T_r$ and $T_d$ the time of the rise and decay of each flare, respectively. Figure \ref{fit_lc} shows the light curve for the main flaring period that occurred in 2011 May, with the fit function superimposed. In Table \ref{fit_par} we reported the fit parameter value.  Using this technique it is also possible to estimate the shortest time variability (to be conservative,  the shortest value extracted is $\Delta t = 2 T_r = 0.33$ days), which is used to put an important constraint on the radiative region size $R_{rad} \leq (c \Delta t\mathcal{D})/(1+z)$, where c is the speed of light, $\mathcal{D}=1/(\Gamma(1-\beta cos\theta))$ is the Doppler beaming factor, $\Gamma$ the bulk Lorentz factor, $\theta$ the viewing angle and $z$ the cosmological redshift. Using the Doppler factors obtained from the SED fitting procedure, which are in agreement with blazars with strong $\gamma$-ray emission \citep{savolainen10},  we can estimate that  $R_{rad}$ is $ \lesssim 2.5 \times 10^{16}$ cm for $\mathcal{D}= 20$ and $ \lesssim 3.8 \times 10^{16}$ cm for $\mathcal{D}= 30$.


\begin{table}
\caption{Parameter values extracted from the flare shape fit.}
\begin{center}
\begin{tabular}{ccc}
\hline
$t_{max}$ &   $T_r \pm err$  & $T_d \pm err$ \\
$(MJD)$  & (day) & (day) \\
\hline
55676.2 &  0.33 $\pm$ 0.07 & 0.81  $\pm$ 0.15 \\
55696.3  & 1.69  $\pm$ 0.24 & 0.29  $\pm$ 0.05 \\
55700.4 & 1.96  $\pm$ 0.34 & 0.54  $\pm$ 0.11 \\
\hline

\label{fit_par}
\end{tabular}
\end{center}

\end{table}

%
%
%
%


%

\section{Radio-to-gamma-ray Spectral Energy Distribution}\label{sect:SED}
%
%

Variability is a powerful diagnostic for the physics of blazars but
creates difficulties in the broad-band SED analysis because
theoretical models can be effectively constrained only with
sufficiently-well time resolved multi-frequency data. For example \fermi-driven observing campaigns
are demonstrating the role of SSC emission for FSRQ objects \citep{bottcher09}, while
the ERC process is being used in fitting also the SEDs of BL Lac objects. First clues, suggesting a smooth transition between the division of blazars into BL Lac objects and FSRQs, are emerging in some studies \citep[e.g., ][]{cavaliere02,giommi12b,sbarrato12}. Our \fermi, \swift\ and \planck\ results on the FSRQ \quattroc\ are pointing out some features more typical of BL Lac objects, and therefore support this hypothesis.



The big blue bump (thermal disc emission) in \quattroc\ appears clearly in the low
emission state SED  (see Figures \ref{fig:SSC} and \ref{fig:EC}), which is based on archival data; however, it is completely
overwhelmed by the continuous synchrotron emission during the $\gamma$-ray flaring
state. The lack of observable thermal disc emission is a common
feature in BL Lac objects, for which contribution from the accretion disc is
negligible in both low and high activity states.


The SED of \quattroc\ obtained during two epochs, flare and post-flare, through \fermi, \swift, \planck\ and radio-optical simultanous observations, appears consistent with a BL Lac object. Even a distinct bulk-Compton spectral excess generated by adiabatic expansion of the emitting region and a cold population of electrons, occasionally observed in some FSRQs, is not evident in the X-ray spectrum of this blazar. 
To evaluate the numerical model of the SED we used the  desktop version  of the  online code developed by A. Tramacere 
\citep[e.g., ][]{massaro06,tramacere09, tramacere11}
which finds  the best-fit parameters of the numerical modelling  by a least-square  $\chi^2$ minimization.

In the following subsections we report simultaneous multi-wavelength SED modeling using both the two-zone SSC and the single zone SSC+ERC scenarios, since these different models both fit the simultaneous SED data of  \quattroc.

The numerical  model self-consistently evaluates the energy content in the resulting equilibrium electron distribution, and compares this value to the magnetic-field energy density \citep{tramacere11}. The minimum energy content of the source is released near equipartition conditions between the magnetic field $B$ power and radiating particle energy power in the jet \citep[e.g., ][]{dermer04,dermer14}. Equipartition ratio values can be used to pick out a preferred scenario because the synchrotron spectrum implies minimum jet power. 
Our SED fit results for \quattroc\ (Fig. \ref{fig:SSC} and  \ref{fig:EC}, Tables \ref{tab:blue}  and \ref{tab:green}) suggest that the single zone, two processes (SSC plus ERC from both disk and torus) scenario is slightly preferred with respect to the two-zone and single process SSC model. In particular the two-zone SSC model fit points out that the energetics are very far from the equipartition condition for the fast emission blob responsible for the inverse Compton GeV $\gamma$-ray component.


%
%

\subsection{Two zones, single SSC process }

\begin{figure*}
 \centering
\hspace{-1cm}
 \includegraphics[width=\hsize]{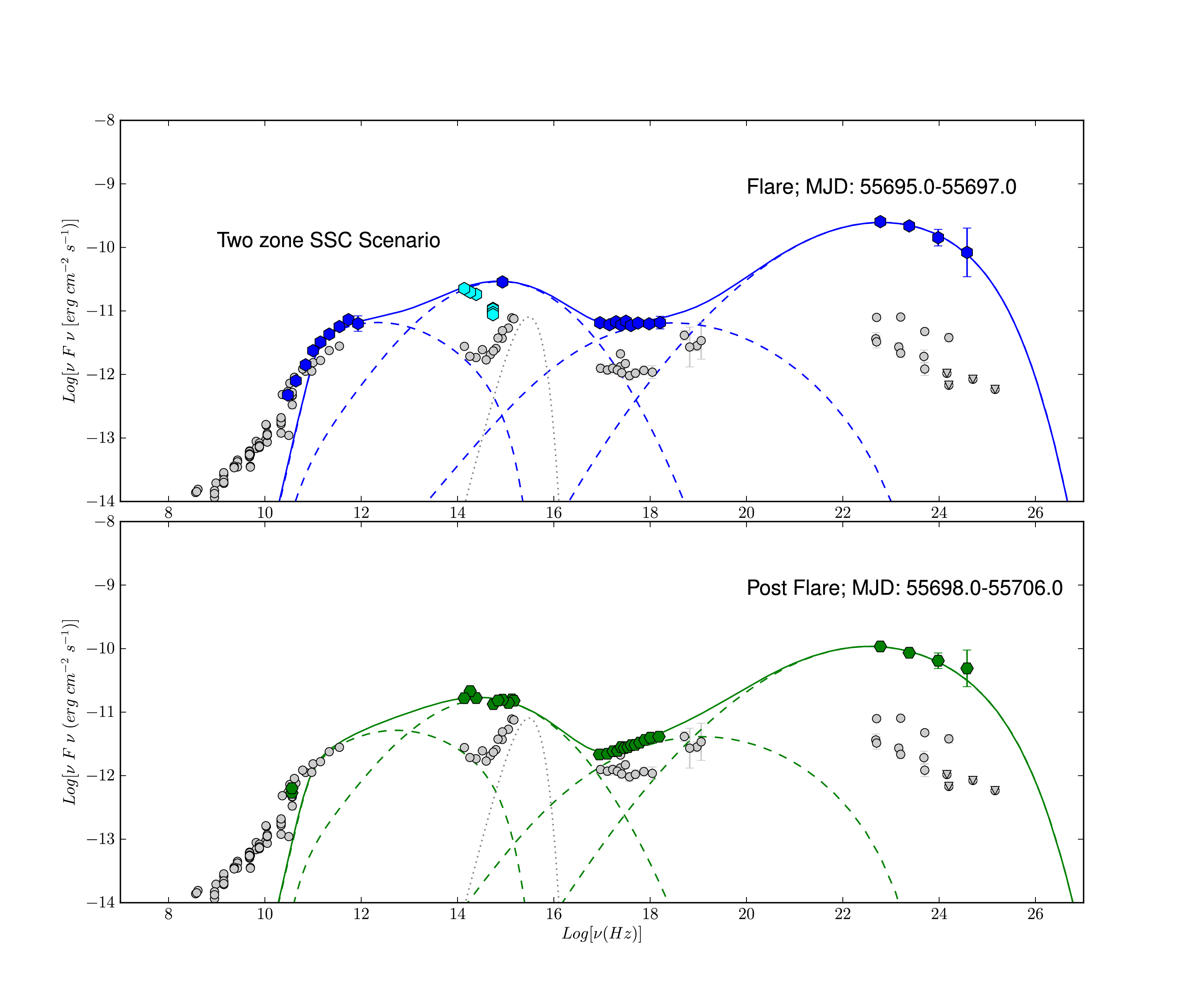}
 \caption{SED of \quattroc\ at different epochs  with fit models (solid lines)  with a two-zone SSC code.  The X- and $\gamma$-ray flare is evident. Simultaneous data during the flare of 2011 May 15  (MJD 55696) are shown in blue; the quasi-simultaneous data of CSS and INAOE are shown in cyan; post flare (i.e. \fermi-LAT and \swift-XRT data are integrated from  2011 May 17 to 25, MJD 55698-55706)  data are shown in green; archival or low state data are shown in grey. The grey dashed lines represent the different components, synchrotron and inverse compton, used in the fitting procedure relative to each zone. The dotted line is the estimated contribution coming from the disc fitted to  archival data.}
  \label{fig:SSC}
\end{figure*}

A single flaring emission zone with a leptonic SSC  process is a model
often used for high-energy peaked (and TeV) blazars. This scenario represents
the first step in  SED modeling attempts. One-zone models usually have difficulty
reproducing highly variable states and composite X-ray spectra often
observed in the SEDs of FSRQs and low/intermediate energy peaked BL
Lac objects. In this view a two-zone SSC model was  applied to
fit the SED of \quattroc , which allows us to take into account, in the fit attempt, the soft X-ray excess and
the XRT spectral shape for the flare state of 2011 May 15 (characterised by a photon index of
2.03 $\pm$ 0.06, Table \ref{swifttab}), and allows us to connect in the fit the simultaneous XRT and LAT data.
Double leptonic emission zones have been recently invoked to fit simultaneous blazar SED data \citep[for example, ][]{georganopoulos03,liu07,latBLLac,tavecchio11,moraitis11}, albeit at the expense of more free parameters.

The flare and post-flare epochs in the SED of \quattroc, based on simultaneous data and on archival data representative of the quiet state, are shown in Figure \ref{fig:SSC} with the two-zone SSC models.

The hypothesis of two regions emitting through the SSC process is recently used in several cases of SED modeling.
For example, a scenario based on a first SSC emission region encompassing the whole jet cross-section plus a second, compact and energetic SSC emission region defined by a high-bulk Lorentz factor blob responsible for the rapidly varying  $\gamma$-ray emission is used in some SED models \citep{tavecchio11,ghisellini08}.

The two emitting blobs for \quattroc\ are thought to represent a compact and faster emission region filled
with fresh and high-energy electrons, and a larger, slower and diluted
region accounting for the radio-band emission from older and
lower-energy cooling electrons, representing the surrounding plasma of the jet. The $\gamma$-ray emission blob is modeled with a bulk Doppler factor  ${\mathcal
  D}_1= 21.1$, size $R_{1}= 7.5\times 10^{15}$ cm, intensity of the tangled magnetic field in the region of $B_{1}= 0.25 $ G.
 The instantaneous electron injection is self-consistently balanced with particle escape on a time scale of the order of $t_{\rm esc} = R/c$.

The radio-band and hard X-ray emitting blob is reproduced using a much larger
emitting region characterised by parameter values  ${\mathcal D}_2=
15$, $R_{2} = 5.5 \times 10^{16}$ cm, and $B_{2}=0.10$ G. Relativistic
Doppler beaming factors of the two zones are found to be consistent
with the range of values found in the maps made from 2008 up to 2012 of the
MOJAVE program  \citep{lister13}. These two regions move relativistically along the jet, oriented at an angle at least $\theta < 38^{\circ}$ with respect to the line of sight.

The kinetic partial differential equation of this model describes the
evolution of the particle energy distribution  after the
injection of freshly accelerated electrons, with an instantaneous rate $Q(\gamma)$ equal to a power
law turning into a log-parabola function in the high-energy tail
\citep{landau86,massaro06} whose functional form is:
\begin{equation}
Q(\gamma)= \left\{ %
\begin{array}{rl}
    (\gamma/\gamma_{0})^{-s} &\mbox{ if  $\gamma\leq\gamma_{0}$ } \\ %
    (\gamma/\gamma_{0})^{-\left(s+r~\log(\gamma/\gamma_{0})\right)} &\mbox{ if $\gamma >
\gamma_{0}$}
\end{array} \right.
\end{equation}
%
where $\gamma_{0}$ is the energy at the turnover frequency, $s$ is the spectral index at the reference energy $\gamma_{0}$ and $r$ is the spectral curvature. The respective values of parameters for
both zones are reported in Table \ref{tab:blue} for the flaring phase
and Table \ref{tab:green} for the post-flare decreasing activity.


\subsection{Single zone, SSC and ERC processes}\label{sect:ssc_erc}

\begin{figure*}
 \centering
\hspace{-1cm}
 \includegraphics[width=\hsize]{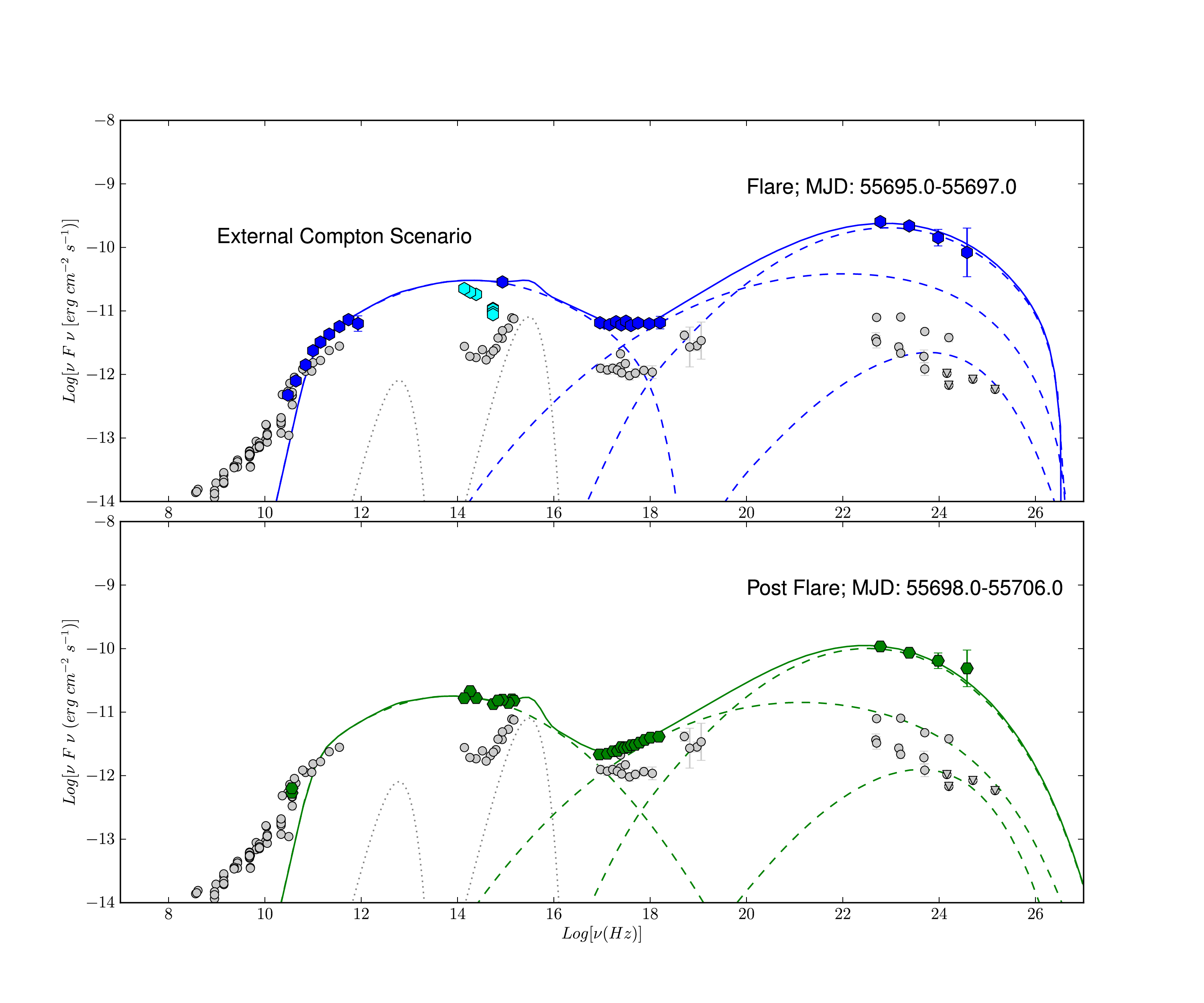}
 \caption{SED of \quattroc\ at different epochs and relative fit models (solid lines) with a single zone and two-processes, SSC and ERC on disc and torus, code. Simultaneous data during the flare of  May 15 are shown in blue; the quasi-simultaneous data of CSS and INAOE are shown in cyan; post flare (i.e. \fermi-LAT and \swift-XRT data are integrated from  May 17 to 25) data are shown in green; archival or low state data are shown in grey.  The grey dashed lines represent the different components, synchrotron and inverse compton, used in the fitting procedure. The dotted line is the estimated contribution coming from the disc  fitted to  archival data and the torus.}
  \label{fig:EC}
\end{figure*}

The high-power MeV-GeV bolometric emission seen in flaring
$\gamma$-ray FSRQs, can usually be better described by external-jet 
Comptonisation of radiation (ERC) models.
In this case the seed photons for the IC process are typically UV photons generated by the accretion disk surrounding the black hole, and reflected toward the jet by the Broad Line Region (BLR) clouds within a typical 
distance from the accretion disk of the order of 1 pc \citep{sikora94}. Another component of external-jet seed photons in the IR band for the scattering is likely provided by a dust torus \citep[DT, see, e.g. ][]{sikora02}. In this case the cooling of relativistic electrons is dominated by Comptonization of near and mid-IR radiation from ambient dust of the torus.
This behaviour has already been found in
FSRQs in the EGRET era
\citep[e.g.,][]{Sokolov05,sikora02, sikora08,sikora09}. A model taking into
account leptonic SSC emission with the relevant addition of leptonic
ERC from a single active blob
(the single zone two-process model) can also be used to explain the
two-epoch SEDs of \quattroc\ as shown in Figure \ref{fig:EC}. An
accretion disc emission component is clearly seen in our archival data
of the quiet activity state (bottom panel of Figure \ref{fig:EC}) and
this could be the origin of the dominant inverse Compton $\gamma$-ray radiation. In
the SSC plus ERC hybrid model for \quattroc\ , X-ray emission can
still be fitted  as an SSC process, while the MeV-GeV $\gamma$-ray
emission detected by \fermi\ can be fit well by dominant ERC
emission from the thermal disc, DT and BLR dissipation region. For this model fit an equilibrium version
of the time-dependent jet model reported in \citet{tramacere11} was
used.

For the ERC emission, all the direct accretion disc radiation field,
accretion disc emission reprocessed in the BLR and the radiation  field from the DT illuminated by the disc are taken into account. We assumed the same electron energy distribution $Q(\gamma)$ with functional form equal to a power-law turning into a
log-parabola function in the high-energy tail described above, for the
single active zone emitting via SSC and ERC processes. In this case
the single emitting blob is modeled with ${\mathcal D}=20.1$, $R= 6.6
\times 10^{16}$ cm and $B= 0.10 $ G. We estimated an accretion disc
with physical characteristics using the UV data.  We used the observations
during the lowest phase of our dataset to constrain the clear sign of
thermal emission coming from the accretion disc and set a reference
value of $L_{disc} \simeq 4 \times 10^{45}$ erg~s$^{-1}$.  The
temperature profile of a standard disc emitting locally as a
black-body following \citet{ghisellini09a} is

\begin{equation}
T_{disc}^4(R)=\frac{3R_SL_{disc}}{16\pi\epsilon\sigma_{SB}R^3}\left(1-  \sqrt{ \frac{3R_s}{R} }\right)
\label{eq:T}
\end{equation}
%
where $\sigma_{SB}$ is the Stefan-Boltzmann constant and
$R_S=2GM_{BH}/$c$^2$ is the Schwarzschild radius. We assumed that the
accretion disc extends from $R_{in}\simeq 3R_S$ to $R_{out}\simeq
500R_s$.  Since $T_{disc}(R)$ peaks at $R\simeq4R_S$ we  used the
 UV observations of the accretion disc to constrain the $T_{disc}$ and use
 Equation \ref{eq:T} to calculate $R_S$.  Using our set of data, we
extracted a value of $T_{disc}\simeq 5 \times 10^4$ K. Assuming an
accretion efficiency $\epsilon\simeq 0.1$, we obtain a $R_s\simeq
1.2\times 10^{14}~$cm. The value of BH mass $M_{BH}$ coming from the
$R_S$ evaluation is about  $4.0 \times 10^{8} M_{\odot}$. This value of $M_{BH}$  we
found using the UV observations of the accretion disc is compatible, assuming an accretion rate
of 0.74 $ M_{\odot}/$yr (coming from the formula that links the
luminosity to the accretion mass rate $L_{disc}=\eta \dot{M}c^2$ erg/s) with
the one obtained by \citet{shields03} using the virial assumption with
the evaluation of FWHM of the broad H$\beta$ line. However it is less than the
 mass of the BH  extracted by \citet{decarli08}  using the
estimation of the luminosity of the host galaxy. The BLR is represented as
a spherical shell of reprocessing material with radial Thomson depth
$\tau_{\rm BLR} = 0.1$. If $L_{disc}=\eta \dot{M} c^2$ erg/s is the total luminosity of the disc (where $\eta$ is the accretion efficiency) the BLR is assumed to be a shell located at a distance from the central black hole of $r_{BLR}= 10^{17} \sqrt{L_{disc}/10^{45}} $ cm \citep{ghisellini08}. The energy dissipation, if it is assumed to be producing the $\gamma$-ray flux, must occur at hundreds of Schwarzschild radii from the black hole, in order to avoid $\gamma \gamma$ pair production absorption.
Similarly the DT is placed at distance from the central black hole of  $r_{\rm DT}= 1 \times 10^{19}$~cm.
%
%
In our model we fixed  values for the distance where the energy dissipation occurs within the BLR region, $r_{\rm BLR}^{min} = 1 \times 10^{18}$~cm and $r_{\rm BLR}^{max} = 2 \times 10^{18}$~cm during the first  flare 2011 May 15 and the post flare phase of 2011 May 17-25. This implies an emission region placed at about  0.32 - 0.65 pc  from the central black hole, in agreement with previous ERC scenarios for FSRQs \citep{dermer09,ghisellini09a}.

Both the two-zone SSC and single zone SSC+ERC models fits appear  appropriate to represent the observed high-state and post-flare SED of \quattroc\ built for the first time with simultaneous data from the three space missions  \fermi, \swift\ and \planck. Equipartition ratio values, numbers of emission components, number of needed model fit parameters, their physical values, the agreements with previous models and VLBA data can alternatively support both  scenarios. The manifest thermal blue-bump  is completely overwhelmed by the non-thermal synchrotron emission during the flare,  evidence which  favours the double-zone SSC picture. Taking into account the multi-epoch VLBA results presented in Section \ref{sect:MOJAVE}  can exclude flaring GeV emission produced by jet knots placed at large distance, of the order of tens of parsecs \citep{lister13}, from the radio core and the BLR region. This can support a more canonical SSC+ERC model for FSRQ sources over the two-zone SSC, because active regions, if located at sub-parsec distances, would lie in proximity of the BLR photon field and if located within a few parsecs would lie at the scale of the molecular dust torus IR photon environment available for Comptonization.


\begin{table}
\caption{SED modeling parameters using a single zone SSC+ERC and  a two zones SSC for the flare period (blue/filled points in Figures \ref{fig:SSC} and \ref{fig:EC}). $^{*}$ indicates that the parameters are frozen to  typical values. $^{**}$ Value obtained describing  the archival data. }
\begin{center}
\begin{tabular}{ccc}
\hline
Parameters & SSC(slow)+SSC(fast)   & SSC+ERC \\
%
\hline
$\chi_{red}^{2}$ & 0.29 & 0.44 \\
\hline
$\log_{10}(R) \textrm{[cm]}$ & 16.74$^{*}$/15.88$\pm$1.06 &  16.82$\pm$0.23\\
B [G]  & 0.10$^{*}$/0.25$\pm$0.04 & 0.10$\pm$0.13 \\
$\mathcal{D}$ & 15$^{*}$/21.1$\pm$0.4 & 20.1$\pm$1.2  \\
$ N [\textrm{cm}^{-3}] $ & 2649.1$\pm$105.3/408.9$\pm$59.9 & 176.9$\pm$19.9 \\
$\log_{10}(\gamma_{min})$ & 0.3$^{*}$/1.9$\pm$0.8  & 1.19$\pm$1.05\\
$\log_{10}(\gamma_{max})$ &  4$^{*}$/6$^{*}$ & 5.50$\pm$1.07\\
\hline
LogParab.+Power-law & & \\
\hline
$ \log_{10}(\gamma_{0})$& 1.78$^{*}$/2.77$\pm$1.26 & 1.83$\pm$0.43 \\
r  & 0.8$^{*}$/1.11$\pm$0.54 & 0.43$\pm$0.05 \\
s & 1.23$\pm$0.04/0.54$\pm$0.12 &  1.32$\pm$0.72\\
\hline
Disc $^{**}$ $-$ BLR $-$ DT & & \\
\hline
$L_{disc}$ [erg/s] &  $-$ &  $4\times 10^{45}$ \\
$T_{disc}^{max}$ [$^\circ $K] & $-$ & $5\times 10^{4}$ \\
$r_{BLR}^{min}$ [cm] & $-$ &  $1\times 10^{18}$ \\
$r_{BLR}^{max}$ [cm] & $-$ &  $2\times 10^{18}$ \\
$\tau_{BLR}$ & $-$ & 0.1 \\
$r_{DT}$ [cm] & $-$ &  $1\times 10^{19}$ \\
$\tau_{DT}$ & $-$ & 0.1 \\
\hline
\end{tabular}
\end{center}
\label{tab:blue}
\end{table}

\begin{table}
\caption{SED modeling parameters using a single zone SSC+ERC and  a two zones SSC for the post-flare period (green/filled points in Figures \ref{fig:SSC} and \ref{fig:EC}). $^{*}$ indicates that the parameters are frozen to a  typical values. $^{**}$ Value obtained describing  the archival data.}
\begin{center}
\begin{tabular}{ccc}
\hline
Parameters & SSC(slow)+SSC(fast)   & SSC+ERC \\
%
\hline
$\chi_{red}^{2}$ & 0.23 & 0.20 \\
\hline
$ \log_{10}(R) \textrm{[cm]}$ & 16.74$^{*}$/15.92$\pm$0.76 & 16.63$\pm$0.77 \\
B [G]  & 0.1$^{*}$/0.07$\pm$0.02 & 0.12$\pm$0.09 \\
$\mathcal{D}$&  15$^{*}$/33.3$\pm$1.9 &  22.9$\pm$5.4 \\
$ N [\textrm{cm}^{-3}] $& 640.6$\pm$22.9/434$\pm$15.3 & 275.2$\pm$16.1 \\
 $ \log_{10}(\gamma_{min})$ & 0.3$^{*}$/0.0009$\pm$0.0001 & 1.74\\
 $ \log_{10}(\gamma_{max})$ &  4.0$^{*}$/6.0$^{*}$ & 10.4$\pm$5.3\\
\hline
LogParab.+Power-law & & \\
\hline
$ \log_{10}(\gamma_{0})$& 1.77$^{*}$/2.77$\pm$0.51 & 1.66$\pm$0.31 \\
r & 0.8$^{*}$/1.04$\pm$0.27 & 0.52$\pm$0.17 \\
s & 0.9$\pm$0.1/0.7$\pm$0.2 & 1.06$\pm$0.15 \\
\hline
Disc $^{**}$ $-$ BLR $-$ DT  & & \\
\hline
$L_{disc}$ [erg/s] &  $-$ &  $4\times 10^{45}$ \\
$T_{disc}^{max}$ [$^\circ $K] & $-$ & $5\times 10^{4}$ \\
$r_{BLR}^{min}$ [cm] & $-$ &  $1\times 10^{18}$ \\
$r_{BLR}^{max}$ [cm] & $-$ &  $2\times 10^{18}$ \\
$\tau_{BLR}$ & $-$ & 0.1 \\
$r_{DT}$ [cm] & $-$ &  $1\times 10^{19}$ \\
$\tau_{DT}$ & $-$ & 0.1 \\
\hline

\end{tabular}
\end{center}
\label{tab:green}
\end{table}

%

%
%

%
%
\section{Discussion and conclusions}\label{sec:discussion}     
%
%
We have presented simultaneous \fermi, \swift\ and \planck, optical and near-IR flux observations and radio-band flux-structure observations of the blazar \quattroc\ (S4~1150+49, OM~484, SBS~1150+497, $z=0.334$). This is one of few cases where time-simultaneous data from three such space missions are available, in particular during a bright GeV flare.
\par 
The GeV outburst of this FSRQ was observed by the LAT around 2011 May 15 after a prolonged period of low $\gamma$-ray activity. The $\gamma$-ray flare was observed simultaneously in  X-ray data, with no measurable time lag (Figure \ref{DCCF}). \quattroc\ showed synchrotron emission peaked in the near-IR and optical wavebands with X-ray spectral softening and time-correlated variability in microwave/X-ray/GeV energy bands as observed more commonly in BL Lac objects rather than FSRQs. As seen in Section \ref{sect:SED} the single SSC mechanism, adopted usually for BL Lac objects,
can also explain the radio-to-gamma-ray SED of this FSRQ during the flaring state and following epoch (Figures \ref{fig:SSC} and \ref{fig:EC}).
\par 
This is also one of the first cases where a two-zone SSC model (slow+fast in-jet components) and a single-zone SSC+ERC model (BLR and DT components)  both appear appropriate to represent the high $\gamma$-ray state multifrequency SED of a FSRQ. Opposite to the majority of the ``strawman'' models overimposed to SED data, our claim follows a true SED modeling through the best-fit model calculations with minimisation over the physical parameter grid \citep{tramacere11}.
\par
We briefly recall the several aspects that make \quattroc\ a particularly  interesting object for high-energy studies. \quattroc\ is a powerful and core-dominated FSRQ showing a bright and structured kiloparsec X-ray jet, diffuse thermal soft X-ray emission produced by the host galaxy and/or galaxy group medium, and a significant fluorescent $K\alpha$ emission line (equivalent width $\simeq 70$ eV) from cold Iron \citep{gambill03,sambruna06b}.
The Fe line detection indicates that even in the X-ray band the beamed jet emission in the low activity states does not completely swamp the accretion-related emission, qualifying this source as a good candidate to investigate the disc-jet connection with multi-frequency observations \citep{grandi04}. This object is also a high luminosity FSRQ characterised by a  one-sided core-jet radio structure, where the strong and  compact jet extends 6 mas (i.e. about 28 pc) in the south-west direction. The 5 GHz VLBI polarisation structure of the source is relatively simple \citep{qi09}, where fractional polarisation ($\sim$1\%) is basically concentrated in the core region, and the direction of the mas-scale magnetic field is consistent with jet direction. \textit{Chandra} resolved and identified in a hot spot of \quattroc\  and compact X-ray substructures \citep{tavecchio05}. Finally simultaneous disc and BLR luminosities show $L_{BLR}/L_{disc}=0.08$ \citep{sambruna06b} and the estimated mass of the SMBH of \quattroc\ is  $4.0\times 10^{8} M_{\odot}$ consistent with what was found by \citet{shields03} and \citet{decarli08}. Our results can be summarised as follows.%

\subsection{A class-transitional FSRQ ?}

\quattroc\ is an FSRQs showing a shift of two orders of magnitude in the frequency of the synchrotron peak (from $\sim10^{12}$ Hz to $\sim10^{14}$ Hz) during the GeV $\gamma$-ray flare. This was accompanied by a contemporaneous marked spectral change in the X-ray energy band. In particular \citet{giommi12} have shown that the distribution of synchrotron emission peaks of FSRQs are centred on a frequency of $10^{13}$ Hz, and the change seen in \quattroc\ can be interpreted as phenomenological transition from a FSRQ to a BL Lac object, with the thermal blue-bump overwhelmed by synchrotron jet emission during the flaring state. This phenomenology can be taken into account in blazar classification and demography paradigms \citep{giommi12b,giommi13} even if occurring in short, transitory, phases of the blazar's life, and can be in agreement with some recent hypotheses suggesting a smooth transition between the division of blazars into BL Lac objects and FSRQs \citep[e.g., ][]{cavaliere02,giommi12b,sbarrato12}. An example of a similar SED peak shift for flaring states is represented by the well-known FSRQ PKS 1510$-$089 \citep{LATpks1510_2010,dammando11}.

\par
 The marked spectral softening of the X-ray spectrum, providing an unusual flat X-ray SED, is also a feature observed in intermediate synchrotron energy peaked BL Lac objects rather than FSRQs. In addition this soft X-ray spectrum does not show any distinct sign of a bulk-Compton origin, generated by the adiabatic expansion of the emitting region and a cold population of electrons, a feature found usually in FSRQs but missing in this case. SED data  therefore suggest a contribution to the X-ray emission  from different emission components, i.e. both synchrotron and inverse Compton (SSC and/or ERC) mechanisms, as usually observed for intermediate energy peaked BL Lac objects \citep[for example ][]{tagliaferri00,ciprini04,abdo2011c}.

\par
Another interesting feature related to this is the optical spectrum. SSDS DR7 and DR8 optical spectra \citep{adelman}, obtained during low emission states, show a rest frame equivalent width (EW) of the broad $H{\alpha}$ emission line of about 300~\AA. We fitted the synchrotron bump of the SED during the $\gamma$-ray flare epoch using a third degree polynomial function and we extrapolated using the best-fit model the value of the continuum at the $H{\alpha}$ frequency. Provided that flux enhancement is due to non-thermal radiation only, the extrapolated continuum of \quattroc\ at the frequency of the $H{\alpha}$ emission line shows an increase of a factor of $\sim 23$ during the $\gamma$-ray flaring state, resulting in a reduction of the emission line EW to about 13\AA, therefore approaching the limit considered for the BL Lac object class (blazars with rest-frame emission line equivalent widths smaller than 5\AA ). The usual classification of blazar subclasses using the rest frame EW definition can be misleading. Objects so far classified as BL Lac objects are turning out to be  two physically different classes: intrinsically weak lined objects, more common in X-ray selected samples, and heavily jet-diluted broad lined sources, more frequent in radio selected samples \citep{giommi12b,giommi13}.

\subsection{Two-zone SSC vs single zone SSC+ERC models}

The multi-frequency SEDs (Figures \ref{fig:SSC} and \ref{fig:EC}) show the synchrotron emission outshining the thermal blue-bump emission that appeared evident in the low activity state. We modeled the radio-to-gamma-ray SEDs for the flare state and the post-flare epoch. A single flaring blob with two different emission mechanisms (SSC and ERC) and a two-zone model with a single SSC process were applied to our SED data. The averaged low state built with archival data (gray points and blue bump signature in Figures \ref{fig:SSC} and \ref{fig:EC}) can be described by including both the jet and disc contributions while the single zone SSC model fails. The disc emission is parametrized in terms of a blackbody from optical to soft X-rays, and the ERC modeling required also a torus emission component contributing to the IC scattering. Our two-zone SSC and single zone SSC+ERC model fits appears both appropriate to represent the observed high-state and post-flare SED of \quattroc, depending on the considered feature (equipartition ratio values, number of parameters, their values, agreements of parameters with previous model values estimated in literature, and agreement with VLBA parameter values and structure). The SSC+ERC is suggested to be slightly more preferred based on equipartition ratio, but the two-zone pure-SSC is  still a valid alternative. The manifest thermal blue-bump  is completely outshone by the non-thermal synchrotron emission during the flare,  evidence which goes in the direction of the pure SSC scenario, that usually better represents the SEDs of BL Lac objects.

\par
On the other hand, the multi-epoch VLBA results presented in Section \ref{sect:MOJAVE} can exclude flaring GeV emission produced by jet knots placed at large distances, of the order of tens of parsecs \citep{lister13}, from the radio core and the BLR region. This suggests a non-negligible contribution of seed photons produced in the BLR for the IC up-scatter, strengthening the case for the hybrid SSC+ERC scenario. 
In general equipartition can be strongly violated during large $\gamma$-ray flare events. Previous SSC models applied to the SED of \quattroc\ and fits of radio-to-X-ray emission of \textit{Chandra}-resolved subcomponents seen in the terminal part of the jet \citep{tavecchio05} support this violation. This suggests IC scattering of synchrotron radiation by some special electron distribution with an excess of high-energy electrons, or CMB photons, or back-scattered central radiation. \fermi-LAT detected  non-spatially resolved GeV emission from \quattroc\ and which may have been an integrated combination of emission from different regions. In this view the $\gamma$-ray flaring state (blue points in Figures \ref{fig:SSC} and \ref{fig:EC})  likely can be better represented by the double zone SSC scenario.

\subsection{Pair production opacity and relativistic beaming}

The $\gamma$-ray flux of \quattroc\ is variable with short time scales ($<$ 1 day). The rapid variability and the large
$\gamma$-ray luminosity imply appreciable pair production opacity. The unbeamed source size estimated from the observed variability timescale indicates that the source is opaque to the photon-photon pair production process if $\gamma$-ray and X-ray photons are produced cospatially. This assumption, however, firmly rests on the simultaneity of the flaring event as observed by \fermi-LAT and \swift-XRT. Simultaneous X-ray and $\gamma$-ray flare events have been measured in the past for other sources like 3C 454.3 \citep{abdo093}. Relativistic beamed jet and emission blobs can solve this problem. Following the arguments given in \citet{mattox93} and adopting
the doubling flux timescale of $t_d \sim 0.6$ days and the observed X-ray
flux of $S_X=(2.1 \pm 0.1)\times10^{-11}~\mathrm{erg}~\mathrm{cm}^{-2}~\mathrm{s}^{-1}$ (as measured during the main flare in $\gamma$ rays) at the observed photon frequency $\nu_X \simeq 10^{18}$ Hz (corresponding to the photons that
interact  with the GeV $\gamma$ rays in the jet rest frame), we can
estimate the Doppler factor $\mathcal{D}$ required for the photon-photon
annihilation optical depth to be $\tau_{\gamma\gamma}<1$. With the
derived relation:
%
\begin{equation}
\tau_{\gamma\gamma} \simeq \frac{\sigma_T d_L^2 S_X}{3t_d c^2 E_X \mathcal{D}^4}
\end{equation}
%
where we assume the emission region linear size $R = c t_d
\mathcal{D}/(1 + z)\simeq 2\times10^{16}$ cm and the source-frame photon energy
$E_X=(1+z)h\nu_x/\mathcal{D}$. Assuming the standard cosmology values we obtain $\mathcal{D} \gtrsim 8.3 $. Omitting the requirement of cospatiality of the X-ray and $\gamma$-ray
emission regions relaxes this limit. This can be compared with the estimate obtained from the VLBA superluminal motion,  $\mathcal{D}>4.3$.
As long as the velocity of the VLBA jet is the same as the
velocity of the outflow within the blazar emission zone, this implies
that the photon-photon annihilation effects involving the X-ray
emission generated within the jet are negligible.

\subsection{Energy dissipation region}

The location of the $\gamma$-ray emitting region is debated, although large distances from the black hole are recently  being favored for about 2/3 of GeV FSRQs \citep[e.g.,][]{marscher10}. Ten epochs of VLBA observations at 15 GHz (MOJAVE program) of \quattroc\ obtained from 2008 May to 2013 February point to Lorentz factor limits that are consistent with our SED modeling and to increasing flux density and polarisation degree in the radio core after the GeV $\gamma$-ray flare. In addition, \planck\ simultaneous observations reveal spectral changes in the sub-mm regime associated with the $\gamma$-ray flare.
Our SED modeling is in agreement with multi-epoch VLBA results and takes into account this evolution observed between the  flare on May 15 (blue SED data and Table \ref{tab:blue}) and the post-flare (May 17-25) epoch (green SED data and Table \ref{tab:green}). The resulting compact emission region of \quattroc\ suggests that nuclear optical/UV seed target photons of the BLR dominate the production of IC emission \citep{tavecchio2010}. Alternatively if the volume involved in the $\gamma$-ray emission is assumed much smaller than jet length  scales, like turbulent plasma cells flowing across standing shocks \citep{marscher2014}, hour/day-scale variability can also be produced at several parsecs from the central engine. The VLBA flux density and polarisation degree in the radio core both increased with the ejection of a new component close in time to the $\gamma$-ray flare epoch.
The jet kinetic power and disc luminosity of \quattroc\ follow the same trend observed for other powerful $\gamma$-ray FSRQs, where a large fraction of the accretion power is converted into bulk kinetic energy of the jet, and our SED models suggest a larger BLR size compared to previous estimates \citep{decarli08,sambruna06b}.

\par $~~~~$  \par

The detailed results about \quattroc\ presented in this work followed the availability of simultaneous \fermi, \swift, \planck\ and VLBA observations triggered by the LAT-detected GeV outburst and by our \swift\ ToO follow-up program. Such data allowed us to investigate multi-frequency flux versus radio-structure relationships, build and constrain pure-SSC vs SSC+ERC SED physical model fits, study emission region localisation, energetics and the evolution of the multi-frequency and high-energy SED during two different emission states for the source, and finally to extrapolate phenomenological features alternatively supporting the FSRQ or BL Lac nature of the source.
 \quattroc\ is a powerful FSRQ, with a FR II morphology and possesses a powerful radio/X-ray jet, but it can have its broad emission lines heavily diluted by a swamping non-thermal continuum during high-energy events. The synchrotron peak energy and the unresolved X-ray spectra  resemble those of intermediate BL Lac objects. Simultaneous multi-frequency data at low and high energies, from space-borne missions like \fermi, \swift, and \planck\ are also needed in the future to correctly draw conclusions about the underlying physics, demography and cosmological evolution of $\gamma$-ray loud AGN.

%
\section*{Acknowledgments}
%
%

We thank Benjamin Walter who helped in the English revision of the paper.
This research has made with the use of the on-line tool for the SED modeling developped by A. Tramacere online at ISDC\footnote{http://www.isdc.unige.ch/sedtool/} and ASDC\footnote{http://tools.asdc.asi.it/SED/}

The \fermi-LAT Collaboration acknowledges generous
ongoing support from a number of agencies and institutes that have supported
both the development and the operation of the LAT as well as scientific data
analysis.  These include the National Aeronautics and Space Administration and the Department of Energy in the United States, the Commissariat \`a l'Energie
Atomique and the Centre National de la Recherche Scientifique / Institut
National de Physique Nucl\'eaire et de Physique des Particules in France, the
Agenzia Spaziale Italiana and the Istituto Nazionale di Fisica Nucleare in
Italy, the Ministry of Education, Culture, Sports, Science and Technology
(MEXT), High Energy Accelerator Research Organization (KEK) and Japan
Aerospace Exploration Agency (JAXA) in Japan, and the K.~A.~Wallenberg
Foundation, the Swedish Research Council and the Swedish National Space Board
in Sweden. Additional support for science analysis during the operations phase is gratefully acknowledged from the Istituto Nazionale di Astrofisica in Italy and the Centre National d'\'Etudes Spatiales in France.

We acknowledge the entire \swift\ mission team for the help and support
and especially the \swift\ Observatory Duty Scientists, ODSs, for their invaluable
help and professional support with the planning and execution of the repeated ToO observations of this target
source. The NASA \swift\ $\gamma$-ray burst Explorer is a MIDEX Gamma Ray Burst mission led by NASA with participation of Italy and the UK.

The \planck\ Collaboration acknowledges the support of: ESA;
CNES and CNRS/INSU-IN2P3-INP (France); ASI, CNR, and INAF (Italy);
NASA and DoE (USA); STFC and UKSA (UK); CSIC, MICINN and JA
(Spain); Tekes, AoF and CSC (Finland); DLR and MPG (Germany); CSA
(Canada); DTU Space (Denmark); SER/SSO (Switzerland); RCN (Norway);
SFI (Ireland); FCT/MCTES (Portugal); and DEISA (EU).

The Mets\"{a}hovi  team
acknowledges the support from the Academy of Finland to our observing
projects (numbers 212656, 210338, and others).

JGN acknowledges financial support from the Spanish
CSIC for a JAE-DOC fellowship, co-funded by the European
Social Fund, and by the Spanish Ministerio de Ciencia e Innovacion,
AYA2012-39475-C02-01, and Consolider-Ingenio 2010, CSD2010-00064, projects.

This research has made use of observations from the MOJAVE database that is maintained by the MOJAVE team. The MOJAVE project is supported under National Science Foundation
grant 0807860-AST and NASA-\fermi\ grant NNX08AV67G. The National
Radio Astronomy Observatory (NRAO VLBA) is a facility of the
National Science Foundation operated under cooperative agreement
by Associated Universities, Inc.

This research has made use of observations from the Catalina Sky Survey, CSS. CSS is funded by the NASA under Grant No. NNG05GF22G issued through the Science
Mission Directorate Near-Earth Objects Observations Program.  The CRTS
survey is supported by the U.S.~National Science Foundation under
grants AST-0909182.

This research has made use of observations obtained with the
2.1~m telescope of the Observatorio Astrof\'{\i}sico Guillermo Haro (OAGH), in the state of Sonora, Mexico, operated by the Instituto Nacional de Astrof\'{\i}sica, \'Optica y Electr\'onica (INAOE), Mexico. OAGH thanks funding from the INAOE Astrophysics Department.

This research has made use of data and software facilities
from the ASI Science Data Center (ASDC), managed by the Italian Space
Agency (ASI).



{\it Facilities:} {\it Fermi}, {\it Swift}, {\it Planck}, {\it INAOE}, {\it CSS}, {\it Mets\"{a}hovi }, {\it MOJAVE}


%

\bsp

\label{lastpage}

\end{document}